\documentclass[paper]{ptephy_v1}

\usepackage{color}
\usepackage{ulem}

\newcommand{\nc}{\newcommand}           
\nc{\vc}[1]     {\mbox{\boldmath $#1$}} 
\nc{\wtil}      {\widetilde}            
\nc{\bras}[1]   {\langle#1|}            
\nc{\kets}[1]   {|#1\rangle}            
\nc{\bra}       {\langle}            
\nc{\ket}       {\rangle}            
\nc{\hO}        {O}           
\nc{\HO}        {\widehat{O}}   

\begin{document}

\title{Complex scaling : physics of unbound light nuclei and perspective}


\author[1,2]{Takayuki Myo}
\affil[1]{General Education, Faculty of Engineering, Osaka Institute of Technology, Osaka 535-8585, Japan
\email{takayuki.myo@oit.ac.jp}
}
\affil[2]{Research Center for Nuclear Physics (RCNP), Osaka University, Ibaraki, Osaka 567-0047, Japan}

\author[3]{Kiyoshi Kat\=o}
\affil[3]{Nuclear Reaction Data Centre, Faculty of Science, Hokkaido University, Sapporo 060-0810, Japan}

\begin{abstract}%
The complex scaling method (CSM) is one of the most powerful methods of describing the resonances 
with complex energy eigenstates, based on non-Hermitian quantum mechanics. 
We present the basic application of CSM to the properties of the unbound phenomena of light nuclei. 
In particular, we focus on many-body resonant and non-resonant continuum states observed in unstable nuclei.
We also investigate the continuum level density (CLD) in the scattering problem in terms of the Green's function with CSM. 
We discuss the explicit effects of resonant and non-resonant contributions in CLD and transition strength functions.
\end{abstract}

\subjectindex{D10, D11, D13}

\maketitle

\section{Introduction}
The physics of resonance is closely related to non-Hermitian physics, because resonances are complex energy states.
At the beginning of the history of quantum mechanics, a complex energy state was introduced by Gamow \cite{Ga28a,Ga28b} to explain the $\alpha$ decay phenomena of nuclei, which emit $\alpha$ particles exponentially, $e^{-\lambda t}$, in terms of time. The real and imaginary parts of the complex energy $E=E_\alpha-i\Gamma_\alpha/2$ describe the energy and decay probability of the emitted $\alpha$ particle, respectively.  He also explained that the decay constant, $\lambda=\Gamma_\alpha/\hbar$, can be evaluated by a tunneling effect of the $\alpha$ particle with the energy $E_\alpha$ for a potential barrier by the repulsive Coulomb force between the $\alpha$ particle and the daughter nucleus. This description of the decay process is based on a time-dependent picture of the wave function. On the other hand, Siegert \cite{Si39} described such a decaying state as a resonant state described by a purely outgoing wave at large distances. The description by Siegert is  based on a stationary-state picture, and then the complex energy state can be obtained as a solution of the time-independent Schr\"odinger equation under the boundary condition for outgoing waves.

Since then, resonances have been extensively investigated in quantum scattering theory \cite{Hu61, La58}, where the real and imaginary parts of the complex energy, corresponding to the resonance energy and width, respectively, are theoretically calculated as a pole of the scattering matrix ($S$-matrix). However, it is difficult for such conventional methods to treat many-body resonances and non-resonant continuum states. Here, we refer to states decaying into more than two-body constituents as ``many-body resonances.''

Recent experimental developments in the field of unstable nuclear physics, starting from the discovery of the neutron halo structure in neutron-rich nuclei such as $^6$He and $^{11}$Li, have shown various interesting phenomena related to the unbound states of nuclei \cite{Ta85,Ta13}. In unstable nuclei, a few extra nucleons are bound to the system with small binding energies, in comparison with the typical value of 8 MeV per nucleon for stable nuclei, and they can be easily released by adding small energies from the outside. This, in turn, implies that the position of the lowest threshold is very close to the ground state and that the coupling effect of the continuum states becomes important even in the ground state. Such a property of unstable nuclei is quite different from that of stable nuclei \cite{Ta96}. One of the interesting features of unstable nuclei is also their so-called Borromean nature of a three-cluster system, in which no two-body subsystem has the bound states. The Borromean nucleus, which has such a feature, has a few bound states and the lowest threshold for a three-body emission, not for a two-body one. Most of the excited states in unstable nuclei are unbound and they are observed as ``many-body resonances.'' A new approach to many-body resonances, therefore, is very desirable in unstable nuclear physics. 

A significant development in the treatment of resonances from two-body systems to many-body systems has been brought about via the complex scaling method (CSM) \cite{Ag71, Ba71, Ho83, Mo98, Mo11}. The wave function of a resonant state with complex energy diverges exponentially at large distances, but such singular behaviour disappears by complex scaling for the resonance wave function.  In CSM, then, annealed resonance wave functions can be described by square-integrable functions in the same way as bound states. Employing an appropriate basis set of square-integrable functions, we can directly obtain many-body resonant states within the eigenvalue problem. As eigenvalue solutions, in addition to resonances, many-body continuum states are obtained on rotated continuum lines (so-called ``$2\theta$'' lines) in the complex energy plane. It is expected, therefore, that CSM can play an important role not only in investigations of the problems of many-body resonances but also in the description of the non-resonant many-body continuum states.

The aim of this article is to provide a brief review of CSM in nuclear physics and its applications to many-body resonant and continuum states of light unstable nuclei. For this purpose, we explain the basic properties of CSM following the studies done by the Hokkaido group \cite{Ao01,Ao06,Ho12,My14}.

In CSM, the many-body Hamiltonian $H$ and wave functions $\Psi$ are transformed by the complex scaling with a parameter $\theta$. We take up several properties of the transformed $H(\theta)$ and $\Psi(\theta)$. The complex-scaled wave functions become non-singular and their matrix elements turn out to be finite, while the original resonance wave functions diverge asymptotically and their matrix elements are infinite. The eigenvalues of $H(\theta)$ are classified into bound, resonant, and continuum states, and these eigenstates are considered to construct a completeness relation (called an extended completeness relation). Continuum states are often discretized and obtained on the $2\theta$ lines. They are, furthermore, classified into various kinds of continuum states, such as two-body and three-body continuum states starting from different threshold energies. Using the eigenvalues and eigenstates, the complex-scaled Green's function can be constructed \cite{Mo98,Mo11_p320}, which is very useful in the calculation of various physical quantities for bound and unbound states, especially concerning the strength functions with respect to the external forces. 

Although the energies of continuum states are discretized with a finite number of basis functions, 
we can obtain a continuum spectrum distribution by projection of their discretized complex energies on the 2$\theta$ lines onto the real energy axis.
By taking appropriate numbers of basis functions for a given $\theta$ parameter, we can confirm that the continuum level density (CLD) is independent of the value of $\theta$. Using CLD, we show that the scattering phase shift can be expressed in a spectral representation. More generally, the Lippmann--Schwinger equation is derived in CSM including many-body scattering contributions explicitly.  This equation in CSM is called the complex-scaled Lippmann--Schwinger (CSLS) equation and is very promising in the study of many-body reaction cross sections.

After explaining these properties of CSM, we show powerful applications to many interesting problems in unstable nuclear physics. In our many applications, we employ Gaussian basis functions \cite{Ka88,Ka89,Hi03} because of their flexibility, simplicity, and high accuracy in many-body problems. In the next section (Sect. 2), the CSM and the framework of our approach are explained. In Sect. 3, illustrative applications of CSM are shown for two-body systems. Further, we present applications to light nuclei as nuclear many-body systems in Sect. 4, and some results using the CSLS equation. in Sect. 5. Finally, a  summary and perspective are given in Sect. 6.

\section{The complex scaling method and the present framework}
\subsection{Complex scaling method}
We briefly explain the complex scaling method proposed by Aguilar, Balslev, and Combes \cite{Ag71,Ba71}. They introduced the transformation (dilation) operator $U(\theta)$ with a scaling parameter $\theta$ for the radial coordinates $\vc{r}_i$ and momentum $\vc{k}_i$ of all particles $i=1,\ldots, n$ (where $n$ is the number of particles) as
\begin{eqnarray}
U(\theta)\vc{r}_iU^{-1}(\theta)=\vc{r}_i\, e^{i\theta},\hspace{1cm}U(\theta)\vc{k}_iU^{-1}(\theta)=\vc{k}_i\,e^{-i\theta},\label{eq-2-1-1}
\end{eqnarray}
where $U(\theta)\,U^{-1}(\theta)=1$. The Schr\"odinger equation for the many-body wave function $\Psi$ is given as
\begin{eqnarray}
H\,\Psi(\vc{r}_1,\ldots,\vc{r}_n)=E\, \Psi(\vc{r}_1,\ldots,\vc{r}_n),\label{eq-2-1-2}
\end{eqnarray}
where the Hamiltonian $H$ subtracting the center-of-mass kinetic energy is defined as 
\begin{eqnarray}
H=\sum_{i=1}^nT_i-T_{\rm c.m.}+\sum_{i<j}^nV(\vc{r}_i-\vc{r}_j),\label{eq-2-1-3}
\end{eqnarray}
which is transformed as 
\begin{eqnarray}
H(\theta)\Psi^\theta=E(\theta)\Psi^\theta,\label{eq-2-1-4}
\end{eqnarray}
where
\begin{eqnarray}
H(\theta)=U(\theta)HU^{-1}(\theta)=e^{-2i\theta}\left[\sum_{i=1}^nT_i-T_{\rm c.m.}\right]+\sum_{i<j}^n V((\vc{r}_i-\vc{r}_j)e^{i\theta}),\label{eq-2-1-5}
\end{eqnarray}
and, for $f=3n-3$,
\begin{eqnarray}
\Psi^\theta=U(\theta)\Psi(\vc{r}_1,\ldots,\vc{r}_n)=e^{if\theta/2}\Psi(\vc{r}_1e^{i\theta},\ldots,\vc{r}_ne^{i\theta}).\label{eq-2-1-6}
\end{eqnarray}
Here we note that $\Psi$ is the internal wave function where the center-of-mass coordinate is separated and excluded, and $f$ is the number of internal degrees of freedom. Important properties of the solutions $\Psi^\theta$ of the complex-scaled Schr\"odinger equation in Eq.~(\ref{eq-2-1-4}) are described in the so-called ABC theorem given by Aguilar, Balslev, and Combes \cite{Ag71,Ba71}.

\begin{figure}[b]
\centering\includegraphics[width=8.5cm]{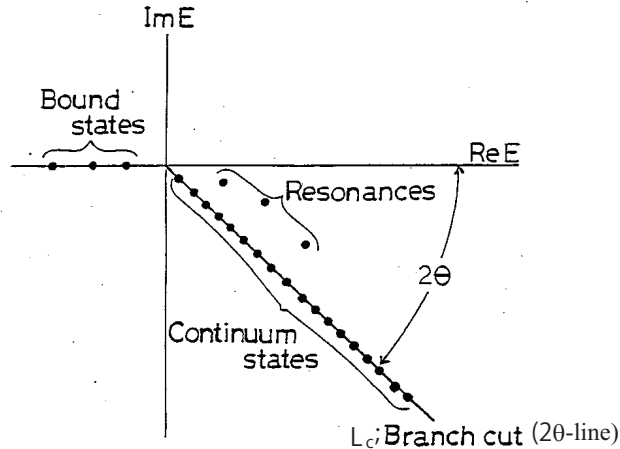}
\caption{Schematic eigenvalue distribution of the complex-scaled Hamiltonian $H(\theta)$ for a single-channel system. Continuum states are discretized on the $2\theta$ line, denoted as solid circles, in the finite basis function method.}
\label{fig-1}
\end{figure}

We solve Eq.~(\ref{eq-2-1-4}) using a basis function method, where the total wave function $\Psi^\theta$ is described with a linear combination of a finite number of the square-integrable basis functions. For simplicity, we explain the method for a single-channel case of a two-particle system. 
The wave function with state index $\alpha$ is expressed as functions of the relative coordinate $\vc{r}$ as
\begin{eqnarray}
\Psi^\theta_\alpha(\vc{r})=\sum_{i=1}^N c_i^\alpha(\theta)\, \bar{u}_i(\vc{r}),\hspace{1.5cm}\langle \bar{u}_i|\bar{u}_j\rangle=\delta_{ij},\label{eq-2-1-7}
\end{eqnarray}
where $\{\bar{u}_i(\vc{r}),~i=1,\ldots,N\}$ is an orthonormal basis set, and the coefficients $c_i^\alpha (\theta)$ ($\alpha=1,\ldots,N$) are obtained by solving the following eigenvalue problem:
\begin{eqnarray}
\sum_{j=1}^N\left\langle \bar{u}_i|H(\theta)|\bar{u}_j\right\rangle c^\alpha_j(\theta)=E_\alpha(\theta)\, c^\alpha_i(\theta).\label{eq-2-1-8}
\end{eqnarray}
The ABC theorem indicates that the eigenvalues are obtained as shown in Fig.~1 and classified as
\begin{eqnarray}
(E_\alpha(\theta),~\Psi^\theta_\alpha)&=&
\renewcommand{\arraystretch}{1.3}
\left\{\begin{array}{lll}
(E_b,~\Psi^\theta_b),& b=1,\ldots,N_b;& \mbox{\underline{b}ound states},\\
(E_r,~\Psi^\theta_r),& r=1,\ldots,N^\theta_r;& \mbox{\underline{r}esonant states},\\
(E_c(\theta),~\Psi^\theta_c),& c=1,\ldots,N-N_b-N^\theta_r;& \mbox{rotated \underline{c}ontinuum states},
\end{array}\right.\label{eq-2-1-9}
\end{eqnarray}
where $N_b$ and $N^\theta_r$ are the numbers of bound and resonant state solutions, respectively, for a given $\theta$ value. The eigenvalues $E_b$ of the bound state solutions, where the number is $N_b$, are negative real values and independent of $\theta$. The complex energies $E_r=E^{\rm res}_r-i\Gamma_r/2$ of the resonant state solutions, which are obtained in the wedge region between the positive energy axis and the $2\theta$ line as shown in Fig.~1, are also independent of $\theta$, but the number $N_r^\theta$ depends on $\theta$ because of the condition $\tan^{-1}(\Gamma_r/2E^{\rm res}_r)<2\theta$ for $E_r=E^{\rm res}_r-i\Gamma_r/2$. The discretized energies $E_c(\theta)$ of continuum states, which are obtained on the $2\theta$ lines, are $\theta$ dependent and expressed as $E_c(\theta)=\epsilon^r_c-i\epsilon^i_c=|E_c(\theta)|\, e^{-2i\theta}$.  

In calculating the matrix elements of the operators $\hat{O}$, it is necessary to take notice of  
the non-Hermiticity of the complex-scaled Hamiltonian $H(\theta)$. For an eigensolution $\Psi^\theta(k)$ of Eq.~(\ref{eq-2-1-4}) with momentum $k$,
we employ its conjugate solution $\widetilde{\Psi}^\theta(k)=\Psi^\theta(-k^*)$ in the biorthogonal state \cite{Mo11,Be68,Ao06,My98}.
This choice is necessary for the eigenstates belonging to the complex energy eigenvalues.
With the biorthogonal solution, the matrix elements are expressed as
\begin{eqnarray}
\bra\widetilde{\Phi}(k)|\hat{O}|\Psi(k')\ket&=&\bra U(\theta)\widetilde{\Phi}(k)|U(\theta)\hat{O}U^{-1}(\theta)|U(\theta)\Psi(k') \ket \nonumber\\
&=&\bra \widetilde{\Phi}^\theta(k) |\hat{O}^\theta|\Psi^\theta(k')\ket, \label{eq-2-1-10}
\\
\hat{O}^\theta&=&U(\theta)\hat{O}U^{-1}(\theta).
\nonumber
\end{eqnarray}
Using the solutions of the eigenvalue problem in  Eq.~(\ref{eq-2-1-8}), the matrix elements are also calculated:
\begin{eqnarray}
\bra \widetilde{\Psi}^\theta_\alpha|\hat{O}^\theta|\Psi^\theta_\beta \ket=\sum_{i,j=1}^N c^\alpha_i(\theta)\, c^\beta_j(\theta)\, \bra \bar{u}_i|\hat{O}^\theta|\bar{u}_j\ket.
\label{eq-2-1-11}
\end{eqnarray}
It is noted that the complex conjugate is not taken for the expansion coefficient $c^\alpha_i(\theta)$, namely the radial part of $\widetilde{\Psi}^\theta_\alpha$
due to the biorthogonal property \cite{Be68,myo97}.

\subsection{Extended completeness relation}
\begin{figure}[h]
\centering\includegraphics[width=15.5cm]{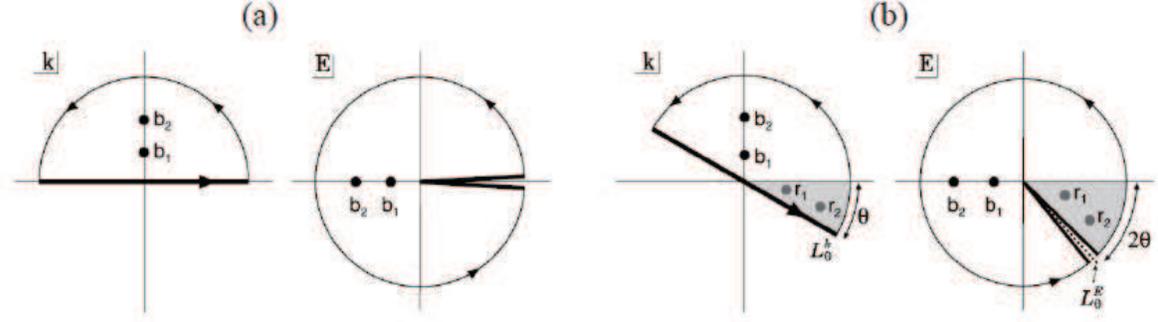}
\caption{Cauchy integral contours in the momentum and energy planes for the completeness relation; (a) without CSM ($\theta=0$) and (b) with CSM ($\theta>0$). The solid circles $b_1$, $b_2$ and $r_1$, $r_2$ are the poles of bound and resonant states, respectively.}
\label{fig-3}
\end{figure}

In standard quantum mechanics without complex scaling, bound and scattering (continuum) states form a complete set that is represented by the completeness relation with real eigenenergies (momenta) of the Hamiltonian $H$ \cite{Ne60} : 
\begin{eqnarray}
\vc{1}&=& \sum_{b=1}^{N_b}|\Psi_b\rangle\langle\Psi_b|+\int_0^\infty dE\,|\Psi_E\rangle\langle\Psi_E|
\nonumber\\
&=&\sum_{b=1}^{N_b}|\Psi_b\rangle\langle\Psi_b|+\int_{-\infty}^\infty dk\,|\Psi_k\rangle\langle\Psi_k|,\label{eq-2-2-1}
\end{eqnarray}
where $\Psi_b$ and $\Psi_E$ ($\Psi_k$) are bound and continuum states, respectively. The continuum states $(\Psi_{-k}, \Psi_k)$ in the momentum representation belong to the states on the real $k$ axis. Therefore,
integration over the real $k$ axis corresponds to that along the rims of the cut of the first Riemann sheet of the energy plane, as shown in Fig. \ref{fig-3} (a).
In the case of a potential problem, the mathematical proof of the completeness relation in Eq. (\ref{eq-2-2-1}) was given by Newton \cite{Ne60} using the Cauchy's theorem.

In CSM, the momentum axis is rotated down by $\theta$, and the poles $(r_1,~r_2, \ldots, r_{N^\theta_r})$ of resonances can enter the semicircle for the Cauchy integration shown in Fig. \ref{fig-3} (b). Then, the resonances are explicitly included in the completeness relation of the complex-scaled Hamiltonian $H(\theta)$ as follows:
\begin{eqnarray}
\vc{1}&=&\sum_{b=1}^{N_b}|\Psi^\theta_b\rangle \langle\widetilde{\Psi}^\theta_b|+
\sum_{r=1}^{N_r^\theta}|\Psi^\theta_r\rangle \langle\widetilde{\Psi}^\theta_r|+
\int_{L_\theta^E}dE\,|\Psi^\theta_E\rangle \langle\widetilde{\Psi}^\theta_E| \nonumber\\
&=&\sum_{b=1}^{N_b}|\Psi^\theta_b\rangle \langle\widetilde{\Psi}^\theta_b|+
\sum_{r=1}^{N_r^\theta}|\Psi^\theta_r\rangle \langle\widetilde{\Psi}^\theta_r|+
\int_{L_\theta^k}dk\,|\Psi^\theta_k\rangle \langle\widetilde{\Psi}^\theta_k|,\label{eq-2-2-2}
\end{eqnarray}
where $\Psi^\theta_b$ and $\Psi^\theta_r$ are the complex-scaled bound and resonant states, respectively. The tilde ($\widetilde{~~}$) in the bra states means the biorthogonal states with respect to the ket states, $\langle\widetilde{\Psi}^\theta_{\alpha}|\Psi^\theta_{\alpha'}\rangle=\delta_{\alpha,\alpha'}$, as mentioned in the previous subsection. Resonant states with number of $N^\theta_r$ are included in the semicircle rotated down by $\theta$ in the momentum plane. The continuum states $\Psi^\theta_E$ and $\Psi^\theta_k$ are the solutions obtained on the rotated cut $L_\theta^E$ ($2\theta$ line) of the Riemann plane and on the rotated momentum axis $L^k_\theta$, respectively. Hereafter, we refer to the relation in Eq.~(13) as the extended completeness relation (ECR) \cite{My98}, which has been proven for single- and coupled-channel systems \cite{Gi03}.

In the case of eigenstates within a finite number of square-integrable basis states, the integration for continuum states is approximated by the summation of discretized states as 
\begin{eqnarray}
\sum_{b=1}^{N_b}|\Psi^\theta_b\rangle \langle\widetilde{\Psi}^\theta_b|+
\sum_{r=1}^{N_r^\theta}|\Psi^\theta_r\rangle \langle\widetilde{\Psi}^\theta_r|+
\sum_{c=1}^{N-N_b-N_r^\theta}|\Psi^\theta_c\rangle \langle\widetilde{\Psi}^\theta_c|~\approx~1.\label{eq-2-2-3}
\end{eqnarray}
Investigation has shown that the reliability of the approximation of the continuum states is confirmed by using a sufficiently large basis number $N$ in CSM \cite{Ao06, My14}.

\subsection{Complex-scaled Green's function}
Application of ECR to the calculations of physical quantities allows us to see contributions from bound, resonant, and continuum states separately. For this purpose, we explain here a spectral expansion of the Green's function in CSM. The complex-scaled expression of the Green's function is 
\begin{eqnarray}
{\cal G}^\theta(E;~\vc{r},~\vc{r}')
&=&
U(\theta)\, {\cal G}(E;~\vc{r},~\vc{r}')\, U^{-1}(\theta)
~=~\left\langle\vc{r}\left|\frac{1}{E-H(\theta)}\right|\vc{r}'\right\rangle,
\label{eq-2-3-1}
\end{eqnarray}
where we used the index $"\theta"$ instead of $(+)$ as the superscript of the outgoing Green's function and dropped the $i\epsilon$ in the denominator.  

\begin{figure}[t]
\centering\includegraphics[width=12.0cm]{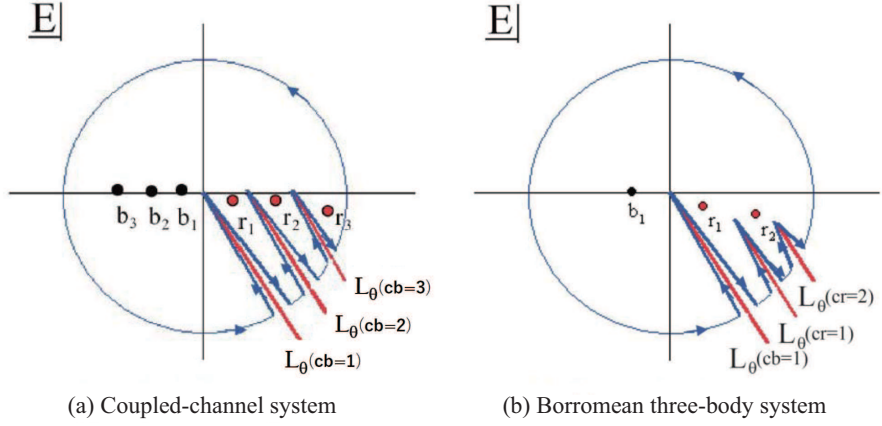}
\caption{The rotated Cauchy integral contours of (a) coupled-channel and (b) Borromean three-body systems on the complex energy plane in CSM.}
\label{fig-4}
\end{figure}

Applying ECR as given in Eq.~(\ref{eq-2-2-2}) to Eq.~(\ref{eq-2-3-1}), we have the complex-scaled Green's function (CSGF)
\begin{eqnarray}
{\cal G}^\theta(E;~\vc{r},~\vc{r}')&=&\sum_{b=1}^{N_b} \frac{|\Psi^\theta_b\rangle \langle\widetilde{\Psi}^\theta_b|}{E-E_b}+
\sum_{r=1}^{N_r^\theta}\frac{|\Psi^\theta_r\rangle \langle\widetilde{\Psi}^\theta_r|}{E-E_r}+
\int_{L_\theta^E}dE_c\, \frac{|\Psi^\theta_{E_c}\rangle \langle\widetilde{\Psi}^\theta_{E_c}|}{E-E_c}.
\label{eq-2-3-2}
\end{eqnarray}
It can be seen that the resonance terms are separated from the continuum one, which constructs a background to the distinguished resonant structure. The continuum term, furthermore, may have various structures in cases of coupled-channel and many-body systems.
In these systems, various kinds of continuum states without CSM may exist degenerately on the real positive energy axis. In CSM, they are rotated and distributed to different channels of the $2\theta$ lines starting from the corresponding threshold energies as shown in Fig.~\ref{fig-4} (a) and (b). The $2\theta$ line of $L_\theta(cb)$ starts from the threshold energies of $E_{cb}$ on the energy axis in the case of coupled-channel systems in Fig.~\ref{fig-4} (a). In many-body systems, as shown in Fig.~\ref{fig-4} (b) for a Borromean three-body case, a $2\theta$ line ($L_\theta(cr)$) may start from the complex energy ($E_{cr}$), sometimes called the resonant threshold, when the subsystem is in resonant states with $E_{cr}$.
   
Thus, the continuum terms of the CSGF for coupled-channel and many-body systems are divided into those on the different $2\theta$ line, and expressed as
\begin{eqnarray}
{\cal G}^\theta(E;~\vc{r},~\vc{r}')&=&\sum_{b=1}^{N_b}\frac{|\Psi^\theta_b\rangle \langle\widetilde{\Psi}^\theta_b|}{E-E_b}+\sum_{r=1}^{N_r^\theta}\frac{|\Psi^\theta_r\rangle \langle\widetilde{\Psi}^\theta_r|}{E-E_r}+\sum_{cb}^{N_{cb}}\int_{L_\theta^{cb}}dE_{cb}\,\frac{|\Psi^\theta_{E_{cb}}\rangle \langle\widetilde{\Psi}^\theta_{E_{cb}}|}{E-E_{cb}}\label{eq-2-3-3}
\end{eqnarray}
and 
\begin{eqnarray}
{\cal G}^\theta(E;~\vc{r},~\vc{r}')&=&\sum_{b=1}^{N_b}\frac{|\Psi^\theta_b\rangle \langle\widetilde{\Psi}^\theta_b|}{E-E_b}+\sum_{r=1}^{N_r^\theta}\frac{|\Psi^\theta_r\rangle \langle\widetilde{\Psi}^\theta_r|}{E-E_r}+\sum_{cb}^{N_{cb}}\int_{L_\theta^{cb}}dE_{cb}\, \frac{|\Psi^\theta_{E_{cb}}\rangle \langle\widetilde{\Psi}^\theta_{E_{cb}}|}{E-E_{cb}}\nonumber\\
& &\hspace{2cm}+\sum_{cr}^{N_{cr}}\int_{L_\theta^{cr}}dE_{cr}\,\frac{|\Psi^\theta_{E_{cr}}\,\rangle \langle\widetilde{\Psi}^\theta_{E_{cr}}|}{E-E_{cr}},\label{eq-2-3-4}
\end{eqnarray}
respectively.

When we solve the complex-scaled Schr\"odinger equation in Eq.~(\ref{eq-2-1-4}) using basis functions of a finite number $N$,
we have the CSGF where the integration in the continuum term is replaced by the sum of the discretized continuum solutions given in Eq.~(\ref{eq-2-2-3})~: for the single-channel case in Eq.~(16),
\begin{eqnarray}
{\cal G}^\theta(E;~\vc{r},~\vc{r}')& \approx &\sum_{b=1}^{N_b} \frac{|\Psi^\theta_b\rangle \langle\widetilde{\Psi}^\theta_b|}{E-E_b}+
\sum_{r=1}^{N_r^\theta}\frac{|\Psi^\theta_r\rangle \langle\widetilde{\Psi}^\theta_r|}{E-E_r}+
\sum_{c=1}^{N-N_b-N_r^\theta}\frac{|\Psi^\theta_{E_c}\rangle \langle\widetilde{\Psi}^\theta_{E_c}|}{E-E_c}.\label{eq-2-3-5}
\end{eqnarray}
In the same way, we have the CSGF for coupled-channel and many-body systems by replacing the integration in Eqs.~(\ref{eq-2-3-3}) and (\ref{eq-2-3-4}), respectively, by the sum of the discretized continuum solutions.

\subsection{Continuum level density}
The density of states at the energy $E$ is defined for the Hamiltonian $H$ as \cite{Le69}
\begin{eqnarray}
\rho(E)={\rm Tr}\,\bigl[\delta(E-H)\bigr].\label{eq-2-4-1}
\end{eqnarray}
From the relation 
\begin{eqnarray}
\frac{1}{E^\pm-H}=P\left[\frac{1}{E-H}\right] \mp i\pi\delta(E-H),\label{eq-2-4-2}
\end{eqnarray}
where $E^\pm=E\pm i\epsilon$, with $\epsilon$ being real and positive and the limit $\epsilon\to 0$ is taken at the final stage, we have 
\begin{eqnarray}
\rho(E)=-\frac{1}{\pi}{\rm Im}\,{\rm Tr}\,\left[\frac{1}{E^+-H}\right].\label{eq-2-4-3}
\end{eqnarray}
The continuum level density (CLD) \cite{Sh92,Kr98,Kr99} is also defined by
\begin{eqnarray}
\Delta(E)=\rho(E)-\rho_0(E)=-\frac{1}{\pi}{\rm Im}\, \left[{\rm Tr}\left\{G(E^+)-G_0(E^+)\right\}\right],\label{eq-2-4-4}
\end{eqnarray}
where the Green's functions are $G(E^+)=1/(E^+-H)$ and $G_0(E^+)=1/(E^+-H_0)$ for the Hamiltonian $H$ and free Hamiltonian $H_0$.  

This CLD describes the density of energy levels which result from the interaction with a finite range, and is related to the scattering $S$-matrix \cite{Le69}:
\begin{eqnarray}
\Delta(E)=\frac{1}{2\pi}{\rm Im}\,\frac{d}{dE}\ln\{\det S(E)\}.\label{eq-2-4-5}
\end{eqnarray}
The scattering $S$-matrix for a single-channel system is expressed as $S(E)=e^{2i\delta(E)}$, where $\delta(E)$ is the scattering phase shift. Then, in the single-channel two-body system, we have
\begin{eqnarray}
\Delta(E)=\frac{1}{\pi}\frac{d\delta}{dE}~~~~~~{\rm and}~~~~~~\delta(E)=\pi\int^{E}_{-\infty}\Delta(E')\,dE'.\label{eq-2-4-6}
\end{eqnarray}
 
Applying CSM to the calculations of Green's functions in CLD, we obtain 
\begin{eqnarray}
\Delta(E)~\approx~\Delta^N_\theta(E)&=&-\frac{1}{\pi}{\rm Im} \left[\, \sum_{b=1}^{N_b}\frac{1}{E-E_b+i\epsilon} +\sum_{r=1}^{N_r^\theta}\frac{1}{E-E_r^{\rm res}+i\Gamma_{r}/2}\right. \nonumber\\
& &\hspace*{1.5cm}\left. +\sum_{c=1}^{N-N_b-N_r}\frac{1}{E-\epsilon^r_c+i\epsilon^i_c} -
                  \sum_{k=1}^{N}\frac{1}{E-\epsilon^{0r}_{k}+i\epsilon^{0i}_{k}}\, \right],\label{eq-2-4-7}
\end{eqnarray}
where $\epsilon^{0r}_{k}-i\epsilon^{0i}_{k}$ ($k=1,\ldots,N$) are the energy eigenvalues of the free Hamiltonian $H_0$. The approximated CLD, $\Delta^N_\theta(E)$, may have a $\theta$ dependence, because a finite number $N$ of basis states are used for diagonalization of the complex-scaled Hamiltonian $H(\theta)$. In the calculation, we adopt a sufficiently large number of $N$ to keep numerical accuracy and to make the $\theta$ dependence negligible in the solutions. Thus, we calculate the phase shift $(\delta(E)\approx\delta^N_\theta(E))$ from $\Delta^N_\theta(E)$:
\begin{eqnarray}
\delta^N_\theta(E)
&=&\int^{E}_{-\infty}dE'\left[\, \sum_{b=1}^{N_b}\pi\,\delta(E-E_b)+\sum_{r=1}^{N_r^\theta}\frac{\Gamma_r/2}{(E'-E_r^{res})^2+\Gamma^2_r/4}\right.\nonumber\\
& &\hspace*{1.5cm}\left. +\sum_{c=1}^{N-N_b-N_r}\frac{\epsilon^i_c}{(E-\epsilon^r_c)^2+(\epsilon^i_c)^2} -\sum_{k=1}^{N}\frac{\epsilon^{0i}_{k}}{(E-\epsilon^{0r}_{k})^2+(\epsilon^{0i}_{k})^2}\, \right].\label{eq-2-4-8}
\end{eqnarray}

By performing the integration for every term, we obtain the spectral decomposition of the phase shift:
\begin{eqnarray}
\delta^N_\theta(E)&=&N_b\, \pi+\sum_{n=1}^{N_r^\theta}\delta_r+\sum_{c=1}^{N-N_b-N_r^\theta}\delta_c-\sum_{k=1}^{N}\delta^0_k,\label{eq-2-4-9}
\end{eqnarray}
where
\begin{eqnarray}
\cot{\delta_r}=\frac{E^{res}_r-E}{\Gamma_r/2},~~~~~\cot{\delta_c}=\frac{\epsilon^r_c-E}{\epsilon^i_c},
~~~~~\cot{\delta^0_k}=\frac{\epsilon^{0r}_k-E}{\epsilon^{0i}_k}.\label{eq-2-4-10}
\end{eqnarray}
 
\subsection{Strength function}
We often investigate the properties of a nucleus by seeing the response to a perturbation of external forces. 
The excitation strength from the initial state to the final state with energy $E$ is expressed in terms of the so-called response function $R_\lambda(E)$ by using the relation of Eq.~(\ref{eq-2-4-2}) as
\begin{eqnarray}
\renewcommand{\arraystretch}{1.9}
\left\{ \begin{array}{lcl}
S_\lambda(E)&=&{\displaystyle\sum_\alpha\bra\widetilde{\Psi}_0|\hat{O}^\dagger_\lambda|\Psi_\alpha\ket\bra\widetilde{\Psi}_\alpha|\hat{O}_\lambda|\Psi_0\ket\, \delta(E-E_\alpha)~=~-\frac{1}{\pi}\mbox{Im}\, R_\lambda(E)},\label{eq-2-5-1}\\
R_\lambda(E)&=&{\displaystyle\int d\vc{r}d\vc{r}'\,\widetilde{\Psi}^*_0(\vc{r})\hat{O}^\dagger_\lambda\, {\cal G}(E;~\vc{r},~\vc{r}')\,\hat{O}_\lambda\Psi_0(\vc{r}')},\label{eq-2-5-2}
\end{array}\right.
\end{eqnarray}
where $E$ is the energy on the real axis, and $\Psi_0$, $\Psi_\alpha$, and $\hat{O}_\lambda$ are the initial and final states and the external force field of rank $\lambda$, respectively.

In CSM, the response function is expressed as 
\begin{eqnarray}
R_\lambda^\theta(E)&=&\int d\vc{r}d\vc{r}'\,\widetilde{\Psi}^{*\,\theta}_0(\vc{r})\hat{O}^{\dagger\,\theta}_\lambda\,{\cal G}^\theta(E;~\vc{r},~\vc{r}')\,\hat{O}^\theta_\lambda\Psi^\theta_0(\vc{r}'),\label{eq-2-5-3}
\end{eqnarray}
for the complex-scaled initial and final states, $\Psi_0^\theta$ and $\Psi_\alpha^\theta$, the Hamiltonian $H(\theta)$ and the external field operator $\hat{O}^\theta_\lambda$. From the decomposition of the CSGF given in Eq. (\ref{eq-2-3-2}), we have the complex-scaled response function in the following decomposed form:
\begin{eqnarray}
R_\lambda^\theta(E)&=&R_{\lambda,B}^\theta(E)+R_{\lambda,R}^\theta(E)+R_{\lambda,C}^\theta(E),\label{eq-2-5-4}
\end{eqnarray}
where
\begin{eqnarray}
\renewcommand{\arraystretch}{2.7}
\left\{ \begin{array}{lcl}
R_{\lambda,B}^\theta(E)&=&{\displaystyle \sum_{b=1}^{N_b}\frac{\bra\widetilde{\Psi}^\theta_0|\hat{O}^{\dagger\,\theta}_\lambda|\Psi_b^\theta\ket\bra\widetilde{\Psi}_b^\theta|\hat{O}_\lambda^\theta|\Psi_0^\theta\ket}{E-E_b}},
 \\
R_{\lambda,R}^\theta(E)&=&{\displaystyle\sum_{r=1}^{N_r^\theta}\frac{\bra\widetilde{\Psi}^\theta_0|\hat{O}^{\dagger\,\theta}_\lambda|\Psi_r^\theta\ket\bra\widetilde{\Psi}_r^\theta|\hat{O}_\lambda^\theta|\Psi_0^\theta\ket}{E-E_r}},
 \\
R_{\lambda,C}^\theta(E)&=&{\displaystyle\int_{L^E_\theta}dE_c(\theta)\frac{\bra\widetilde{\Psi}^\theta_0|\hat{O}^{\dagger\,\theta}_\lambda|\Psi_{E_c}^\theta\ket\bra\widetilde{\Psi}_{E_c}^\theta|\hat{O}_\lambda^\theta|\Psi_0^\theta\ket}{E-E_c(\theta)}}.\end{array}\right.
\end{eqnarray}
Applying this decomposition of the response function, the complex-scaled strength function, $\displaystyle S_\lambda^\theta(E)=-\frac{1}{\pi}$Im\,$R_\lambda^\theta(E)$, is also decomposed as 
\begin{eqnarray}
S_\lambda^\theta(E)&=& S_{\lambda,B}^\theta(E)+S_{\lambda,R}^\theta(E)+S_{\lambda,C}^\theta(E).
\end{eqnarray}

The matrix elements of the complex-scaled operator given in Eq.~(\ref{eq-2-1-10}) are independent of $\theta$ \cite{Ho97}. While $R^\theta_{\lambda,B}$ for bound states are $\theta$ independent, the $\theta$ dependence of $R^\theta_{\lambda,R}$ and $R^\theta_{\lambda,C}$ originates from $N_r^\theta$, $L^E_\theta$, and $E_c(\theta)$. 
However, the final result of the strength function $S_\lambda^\theta(E)$ is an observable, which is positive definite and independent of $\theta$, i.e. $S_\lambda^\theta(E)=S_\lambda(E)$, though each term of $S_{\lambda,R}^\theta(E)$ and $S_{\lambda,C}^\theta(E)$ may have negative numbers. In coupled-channel and many-body systems, the continuum-state term of the CSGF is decomposed further into many kinds of continuum states on the different $2\theta$ lines. Owing to such a decomposition of the final-state contributions, we can investigate which state largely contributes to the formation of the structures observed in the strength function. This is a prominent feature of CSM and is applicable to many-body unbound states such as the Borromean three-body systems, as shown in Sect. 4.   

\subsection{Complex-scaled Lippmann--Schwinger equation}
We proposed a method of describing many-body scattering using CSM, referred to as the complex-scaled solutions of the Lippmann--Schwinger equation (CSLS) \cite{Ki09,Ki10,Ki11,Ki13}. In the CSLS method, the scattering states are described by combining the formal solutions of the Lippmann-Schwinger equation with the CSGF defined by Eq.~(\ref{eq-2-3-1}), which automatically satisfies the correct boundary conditions using the complex-scaled eigenstates.

The formal solution of the Lippmann--Schwinger equation is expressed as
\begin{eqnarray}
\Psi^{(+)}=\Phi_0+\frac{1}{E^+-H}V\Phi_0,\label{eq-2-6-1}
\end{eqnarray}
where $\Phi_0$ is a solution of an asymptotic Hamiltonian $H_0$. The total Hamiltonian is also given as $H=H_0+V$.

In the CSLS method, we utilize the CSGF given in Eq.~(\ref{eq-2-3-1}), which is related to the non-scaled Green's function ${\cal G}(E,\vc{r},\vc{r}')$ with outgoing boundary conditions as
\begin{eqnarray}
\frac{1}{E^+-H}&=&{\cal G}(E,\vc{r},\vc{r}')~=~U^{-1}(\theta)\,{\cal G}^\theta(E,\vc{r},\vc{r}')\,U(\theta).\label{eq-2-6-2}
\end{eqnarray}
Using the eigenstates $\Psi^\theta_\nu$ of the complex-scaled Hamiltonian $H(\theta)$ with state index $\nu$ and their biorthogonal states $\widetilde{\Psi}^\theta_\nu$, we rewrite the Green's function in Eq.~(\ref{eq-2-6-2}):
\begin{eqnarray}
{\cal G}(E,\vc{r},\vc{r}')=\sum_\nu\hspace{-0.6cm}\int~~U^{-1}(\theta)|\Psi^\theta_\nu\ket\frac{1}{E-E^\theta_\nu}\bra\widetilde{\Psi}^\theta_\nu|U(\theta).\label{eq-2-6-3}
\end{eqnarray}
Thus we have the outgoing scattering state as
\begin{eqnarray}
|\Psi^{(+)}\ket=|\Phi_0\ket+\sum_\nu\hspace{-0.6cm}\int~~U^{-1}(\theta)|\Psi^\theta_\nu\ket\frac{1}{E-E^\theta_\nu}\bra\widetilde{\Psi}^\theta_\nu|U(\theta)V|\Phi_0\ket. \label{eq-2-6-4}
\end{eqnarray}
 
The scattering amplitude $f(k)$ of a two-particle collision with energy $\displaystyle E=\frac{\hbar^2k^2}{2\mu}$ is calculated from the following expression \cite{Ta72, Kr07}:
\begin{eqnarray}
f(k)=-\frac{2\mu}{\hbar^2k^2}\bra \Phi_0(k)|V|\Psi^{(+)}(k)\ket=f^{\rm Born}(k)+f^{\rm sc}(k),\label{eq-2-6-5}
\end{eqnarray}
where
\begin{eqnarray}
\renewcommand{\arraystretch}{2.0}
\left\{ \begin{array}{lcl}
f^{\rm Born}(k)&=&{\displaystyle-\frac{2\mu}{\hbar^2k^2}\bra \Phi_0(k)|V|\Phi_0(k)\ket}\label{eq-2-6-6}
\\
f^{\rm sc}(k)&=&{\displaystyle-\frac{2\mu}{\hbar^2k^2}\sum_\nu\hspace{-0.6cm}\int~~\bra \Phi_0(k)|V|U^{-1}(\theta)|\Psi^\theta_\nu\ket\frac{1}{E-E^\theta_\nu}\bra\widetilde{\Psi}^\theta_\nu|U(\theta)V|\Phi_0(k)\ket}\label{eq-2-6-7}\end{array}\right.
\end{eqnarray}
The first term is the Born term and the second is a spectrum decomposition of the scattering amplitude for a short-range interaction $V$. In the case of a long-range interaction such as the Coulomb potential, it should be included in the Hamiltonian $H_0$ and $\Psi_0$ is expressed by using the asymptotic Coulomb wave functions. 

In the coupled-channel case of two-body initial and final systems, the transition amplitude from $\alpha$ to $\beta$ channels is 
\begin{eqnarray}
f_{\beta\alpha}(k_\beta,k_\alpha)=-\frac{2\sqrt{\mu_\alpha\mu_\beta}}{\hbar^2k_\alpha k_\beta}\bra \Phi_0(k_\alpha)|V|\Psi^{(+)}_\beta(k_\beta) \ket=f^{\rm Born}_{\beta\alpha}(k_\beta,k_\alpha)+f^{\rm sc}_{\beta\alpha}(k_\beta,k_\alpha),\label{eq-2-6-8}
\end{eqnarray}
and
\begin{eqnarray}
\renewcommand{\arraystretch}{2.2}
\left\{ \begin{array}{lcl}
f^{\rm Born}_{\beta\alpha}(k_\beta,k_\alpha)&=&{\displaystyle-\frac{2\sqrt{\mu_\alpha\mu_\beta}}{\hbar^2k_\alpha k_\beta}
\bra \Phi_{0}(k_\alpha)|V|\Psi_{0}(k_\beta)\ket}   \label{eq-2-6-9}\\
f^{\rm sc}_{\beta\alpha}(k_\beta,k_\alpha)&=&{\displaystyle-\frac{2\sqrt{\mu_\alpha\mu_\beta}}{\hbar^2k_\alpha k_\beta}\sum_\nu\hspace{-0.6cm}\int~~\bra \Phi_{0}(k_\alpha)|V|U^{-1}(\theta)|\Psi^\theta_\nu\ket\frac{1}{E-E^\theta_\nu}\bra\widetilde{\Psi}^\theta_\nu|U(\theta)V|\Phi_0(k_\beta) \ket,} \label{eq-2-6-10}\end{array}\right.
\end{eqnarray}
where $V$ is a coupling interaction.

\section{Illustrative application to simple two-body systems}
In this section we explain the characteristics of the CSM approach by showing the application of the 
present framework explained in Sect. 2 to simple two-body systems. 

\subsection{Schematic potential model}
\begin{figure}[b]
\begin{center}
\includegraphics[width=6.2cm,clip]{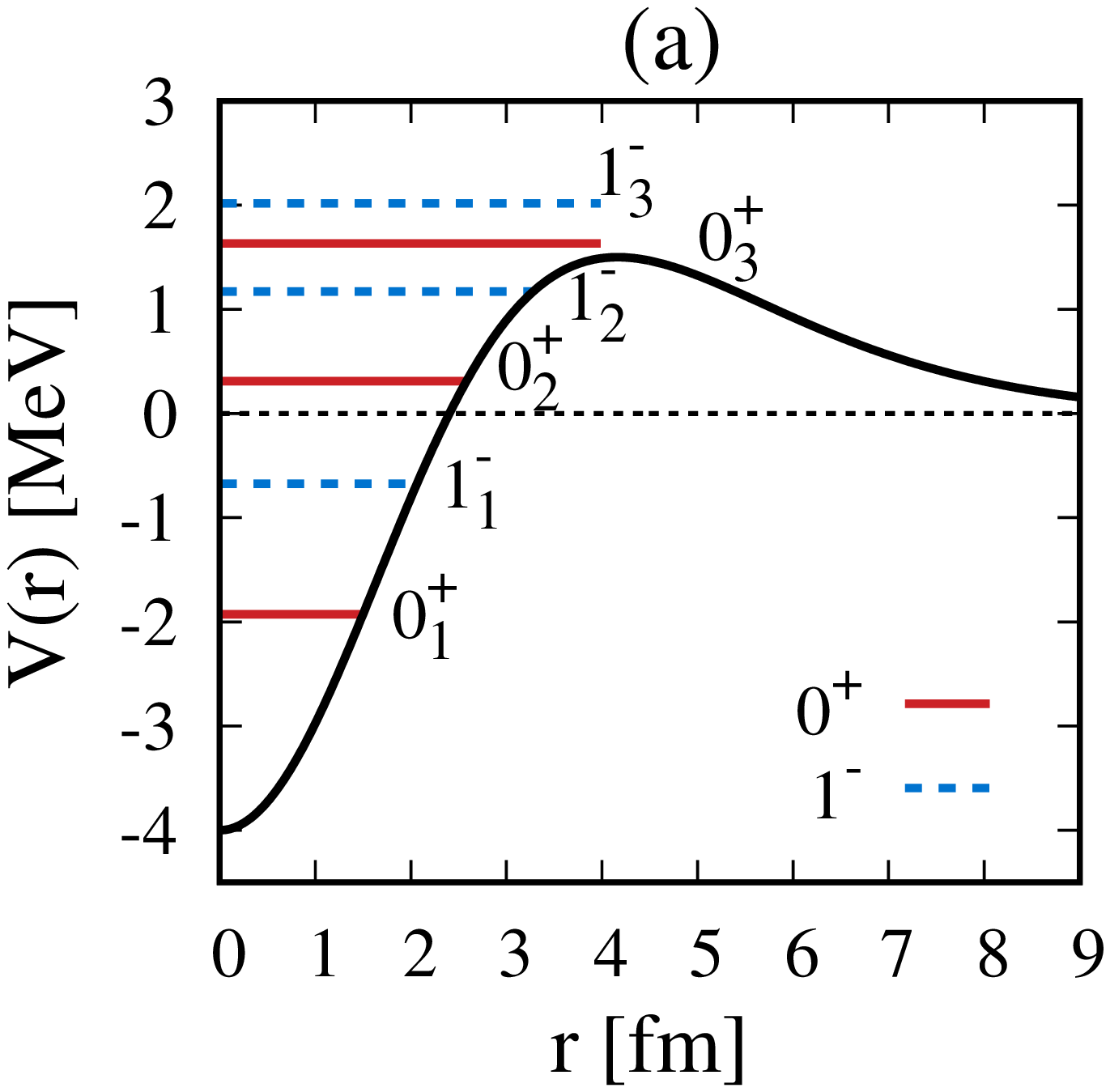}~~~
\includegraphics[width=8.7cm,clip]{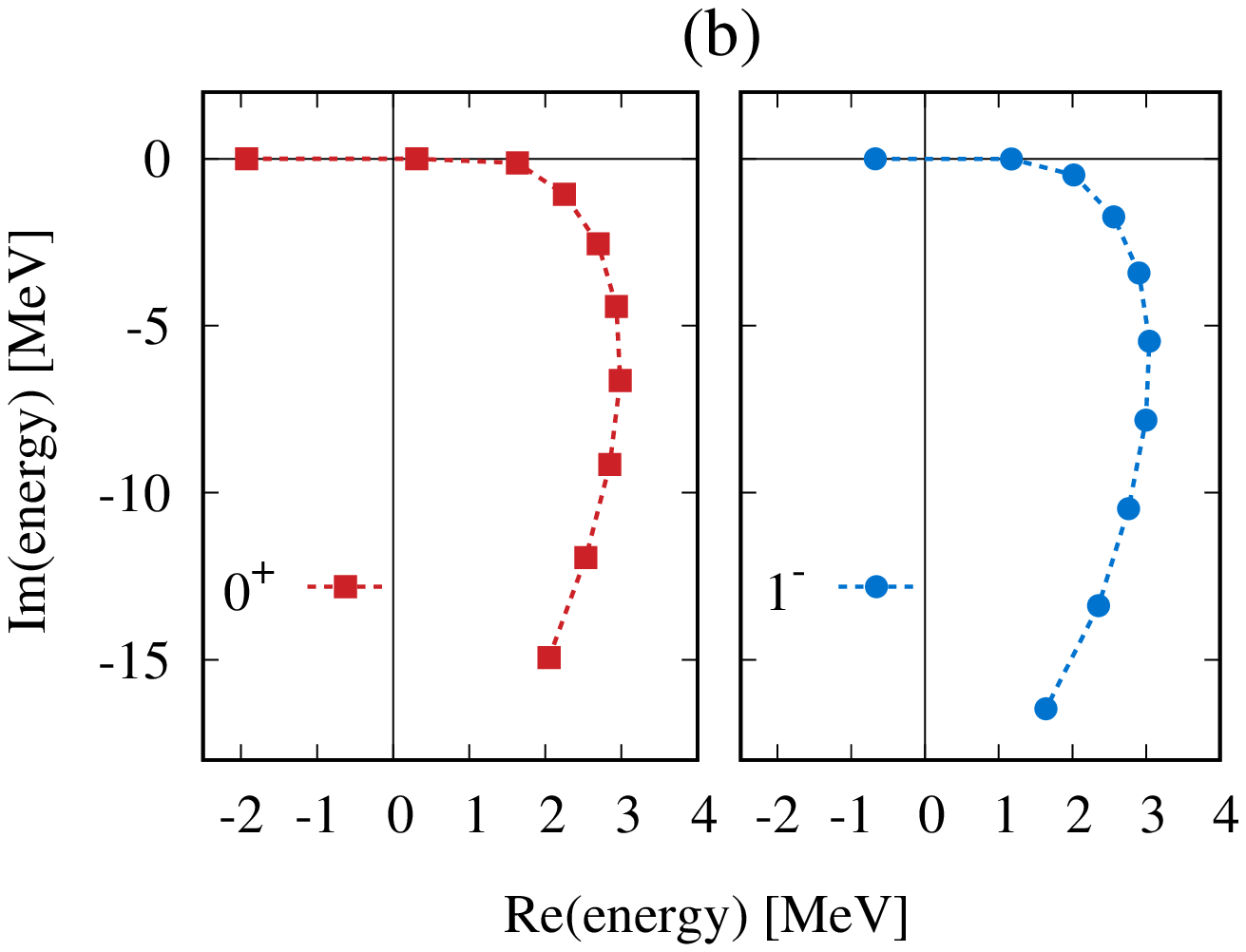}
\caption{a) Potential shape and low-lying energy levels for $0^+_1$ and $1^-$ states. 
b) Energy eigenvalues in the complex energy plane.}
\label{fig:G-pot}
\end{center}
\end{figure}

We demonstrate the calculation of resonances using CSM.
For this purpose, we employ the potential model \cite{Ho97,My98,csoto90,myo97} with the simple Hamiltonian given as
\begin{eqnarray}
H=-\frac{1}{2}\nabla^2+V(r),
\hspace*{0.8cm}
V(r)=-8e^{-0.16r^2}+4e^{-0.04r^2}.
\label{eq:csoto_pot}
\end{eqnarray}
In this model, we calculate the $J^\pi=0^+$ and 1$^-$ states and also the dipole strength function in CSM.
The energy spectra are shown in Fig.~\ref{fig:G-pot}; panel (a) shows the several energy levels with the potential shape
and (b) is the energy eigenvalue distribution in the complex energy plane. 
There is one bound state in each state and we obtain many resonances including the ones with large decay widths in CSM.

\begin{figure}[bh]
\centering
\includegraphics[width=7.5cm,clip]{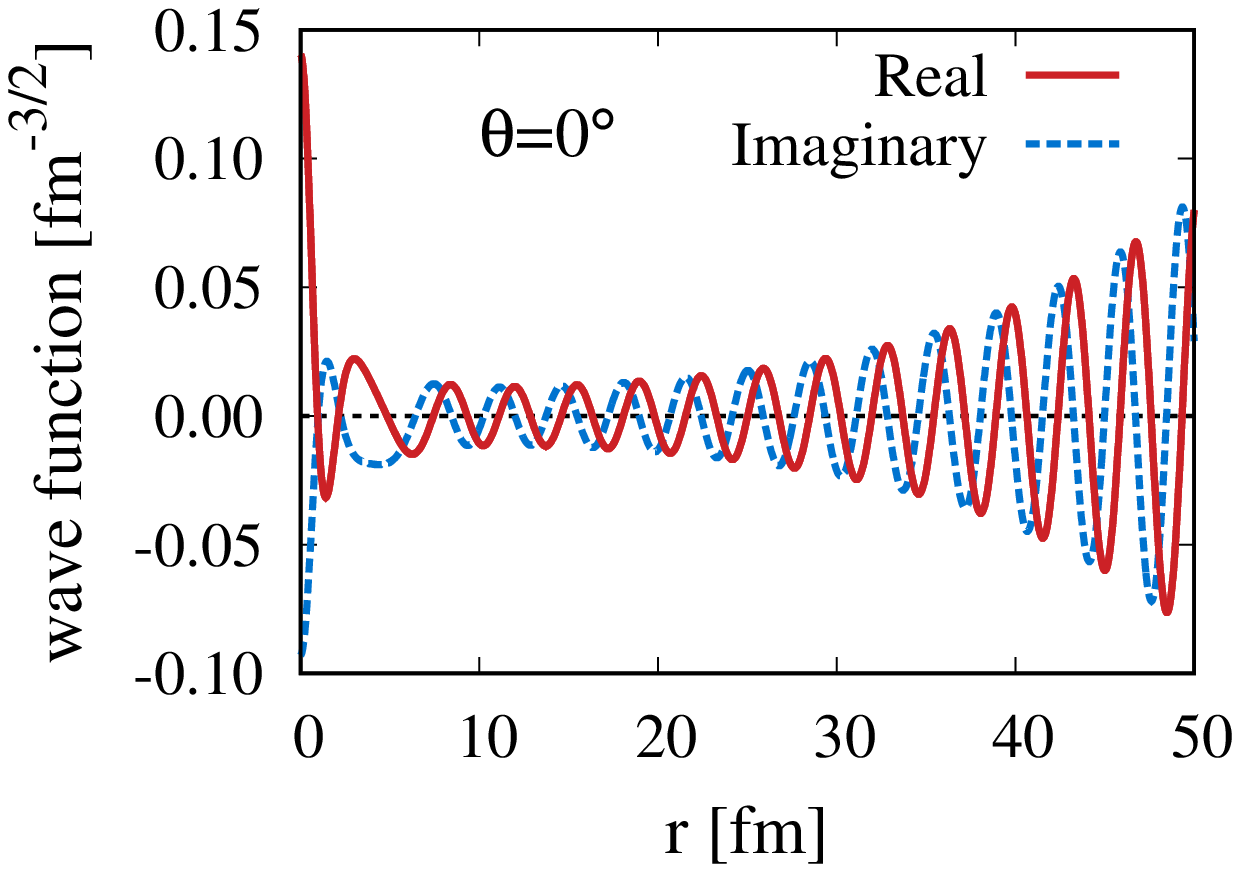}~~~
\includegraphics[width=7.5cm,clip]{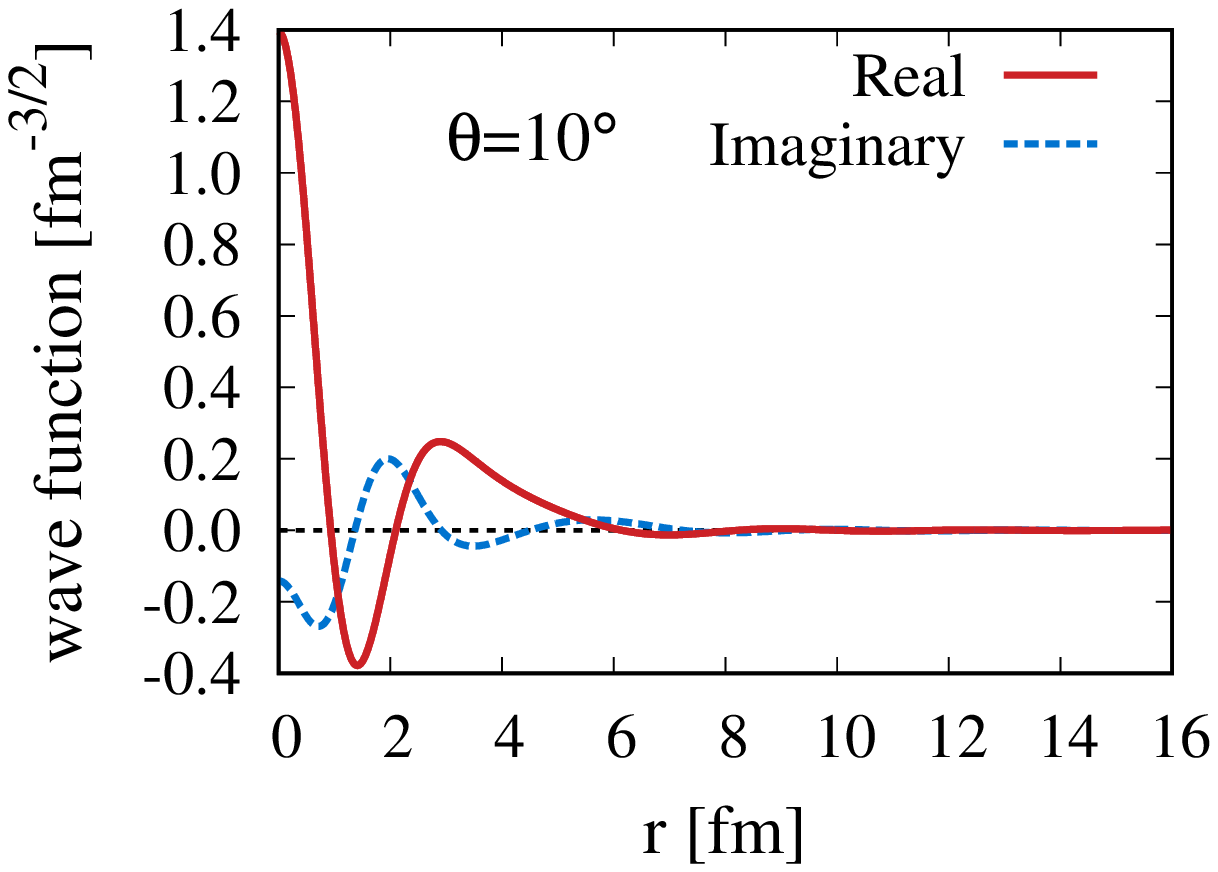}
\caption{Wave functions of the $0^+_3$ resonant state in the schematic potential model; left: without CSM, and right: with CSM.}
\label{fig:WF}
\end{figure}

In Fig. \ref{fig:WF} we show the wave function of the $0^+_3$ state, 
the resonance energy of which is located at the top of the potential barrier
($E_r=1.63$ MeV, $\Gamma=0.246$ MeV), as shown in Fig.~\ref{fig:G-pot} a).
The left panel of Fig. \ref{fig:WF} shows the wave function without CSM, i.e. obtained directly under the divergent boundary condition at the asymptotic region. 
The wave function is complex and its divergent behavior is clearly confirmed.
The right panel of Fig. \ref{fig:WF} shows the wave function transformed in CSM with a scaling angle of $10^\circ$.
The damping behavior of the wave function is confirmed around 10 fm in CSM.


\begin{figure}[t]
\centering
\includegraphics[width=15.0cm,clip]{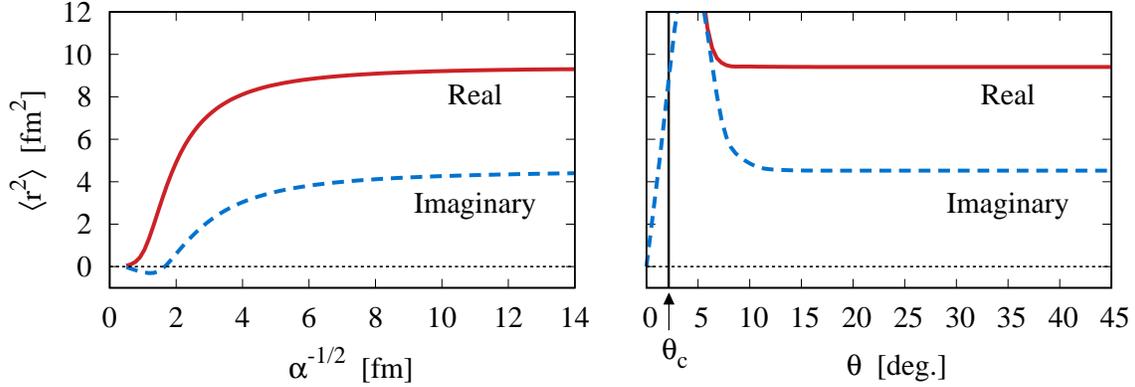}
\caption{Matrix element of $r^2$ for $0^+_3$ in the schematic potential model.
Left: Convergence factor method with the parameter $\alpha$ \cite{zeldovich61}. 
Right: Complex scaling method with the scaling angle $\theta$ \cite{Ho97}.
The arrow indicates the critical angle $\theta_{\rm c}$ for the $0^+_3$ resonance.}
\label{fig:radius}
\end{figure}

We discuss the matrix elements of resonances using the wave functions obtained in CSM.
In Fig. \ref{fig:radius} we show the calculation of the matrix element of $r^2$ for the $0^+_3$ resonant state in two different ways.
One is that the resonance wave function without CSM is used and the convergence factor is introduced
in the numerical integration of the matrix element \cite{zeldovich61,romo68,berggren70,garmati71}.
In general, the resonance wave functions diverge exponentially at an asymptotic region
and the matrix elements of resonances are difficult to calculate directly in a usual integration.
This property has been discussed as being evaluated by introducing a convergence factor such as $e^{-\alpha r^2}$ and
taking continuation to the limit of $\alpha \rightarrow 0$ \cite{zeldovich61}:
\begin{eqnarray}
  \langle\widetilde{\Phi}|\widehat{O}|\Phi\rangle
  &=&\lim_{\alpha \rightarrow 0} \int \widetilde\Phi^*(\vc{r})\, \widehat{O}\, \Phi(\vc{r})\, e^{-\alpha r^2} d\vc{r},
\end{eqnarray}
where $\widehat{O}$ is an arbitrary operator acting on the resonance wave function $\Phi$.

In CSM, the above matrix element is transformed using the complex-scaled wave function $\Phi^\theta=U(\theta)\Phi$ and
the corresponding complex-scaled operator $\widehat{O}^\theta=U(\theta)\widehat{O}U^{-1}(\theta)$ as
\begin{eqnarray}
  \langle\widetilde{\Phi}|\widehat{O}|\Phi\rangle
&=& \langle\widetilde{\Phi}^\theta|\widehat{O}^\theta|\Phi^\theta\rangle.
\end{eqnarray}

We show the matrix element of $\widehat{O}=r^2$ for the $0^+_3$ resonance in a schematic potential case in Fig. \ref{fig:radius}
, which becomes a complex number.
We display two kinds of calculation of the squared radius of the $0^+_3$ resonance, 
with $\alpha$ dependence in the convergence factor method \cite{zeldovich61} and $\theta$ dependence in CSM. 
In CSM, we can see that both real and imaginary values of 
the squared radius become independent of the scaling parameter 
$\theta$ after passing the critical angle $\displaystyle \theta_{\rm c}=\frac{1}{2} \tan^{-1}(\Gamma/2E_r)$. 
In practical calculations, we use a finite number of basis functions
and then have to take a $\theta$ value rather larger than $\theta_{\rm c}$. 
We can see good agreement between the results of CSM and the convergence factor method for both real and imaginary parts.
In CSM, when we take a sufficient value of $\theta$, we can discuss the properties of resonances from their matrix elements.

\begin{figure}[t] 
\centering
\includegraphics[width=10.0cm,clip]{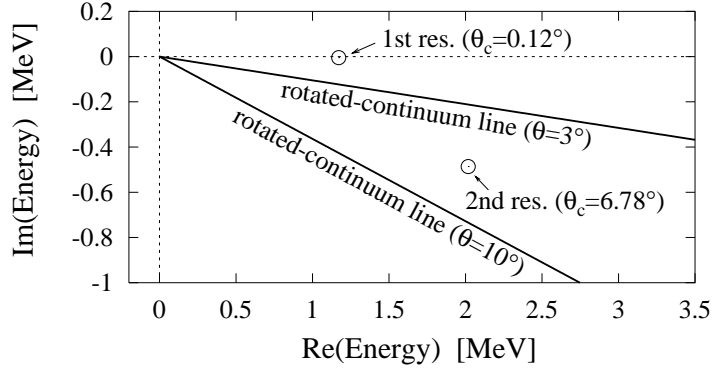}
\caption{Energy eigenvalues of the resonances of the first and second $1^-$ states in the complex energy plane with two rotated continuum lines at $\theta=3^\circ$ and 10$^\circ$.}
\label{fig:csoto-pole}
\end{figure}

In CSM, we obtain the CSGF using the complex-scaled wave functions not only of resonances but also non-resonant continuum states. This Green's function is utilized in the calculation of various kinds of strength functions, 
and also the few-body scattering problems.
We take the example of the calculation of the dipole strength function from $0^+_1$ to the scattering states of $1^-$ in CSM,
where the transition operator is given as $\widehat{O}=rY_{10}(\hat{r})/\sqrt{4\pi}$ \cite{My98}.
We discuss the effects of resonances and non-resonant continuum states of $1^-$ states on the dipole strength.
For $1^-$ states, when the scaling angle $\theta$ is 3$^\circ$ we obtain one bound state ( $E_{0}$=$-$0.68 MeV ), 
one resonance ( $E_1=1.17-i0.49 \times 10^{-2}$ MeV ), and the residual non-resonant continuum states.
From Fig.~\ref{fig:csoto-pole}, the second resonance is not obtained for small $\theta$.
When $\theta$ is $10^\circ$, one new resonance ( $E_2=2.02-i0.49$ MeV ) is decomposed from the continuum states.

\begin{figure}[t]
\centering
\includegraphics[width=12.0cm,clip]{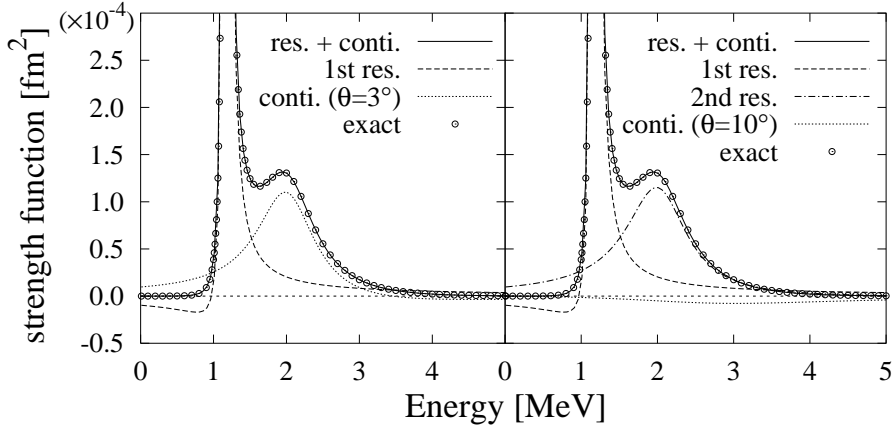}
\caption{Dipole transition strength in the schematic potential model from the $0^+$ ground state to the $1^-$ states with CSM using $\theta=3^\circ$ (left) and 10$^\circ$ (right).
The dashed and dash-dotted lines are the contributions from the first and second resonances, respectively. 
The dotted lines are the continuum contributions, and the solid lines are the sum of all the terms. 
The open circles are the exact calculation by solving the scattering problem for $1^-$ states.}
\label{fig:csoto-E1}
\end{figure}

In Fig.~\ref{fig:csoto-E1}, we show the dipole strength function from the $0^+$ ground state into the 1$^-$ states. 
We compare the strengths obtained in CSM with the exact calculation (open circles), 
which is obtained by solving the ordinary scattering problem for $1^-$ states.
We calculate the distribution using the CSGF and the matrix elements of the dipole transition in CSM, 
which is shown in the solid line and agrees with the exact one.
In the figure, a sharp peak is observed just above 1 MeV. 
The dashed curve denotes the contribution from the first resonance. 
We conclude that the sharp peak originates from the first resonance. 
The left panel of Fig.~\ref{fig:csoto-E1} is the results with a small $\theta$ of $3^\circ$, and shows that 
the continuum contribution (dotted curve) forms a second peak around 2 MeV. 
The right panel of Fig.~\ref{fig:csoto-E1} shows the case in which a large $\theta$ of 10$^\circ$ is used to describe the second resonance. 
It is found that the contribution of the second resonance (dash-dotted curve) is a major component in making a peak around 2 MeV, 
whereas the contribution of the residual continuum states (dotted curve) is very small. 
From these results, it is concluded that two two major peaks of dipole strength come from the contributions of two 1$^-$ resonances.

\begin{figure}[t]
  \begin{center}
    \includegraphics[width=10.0cm]{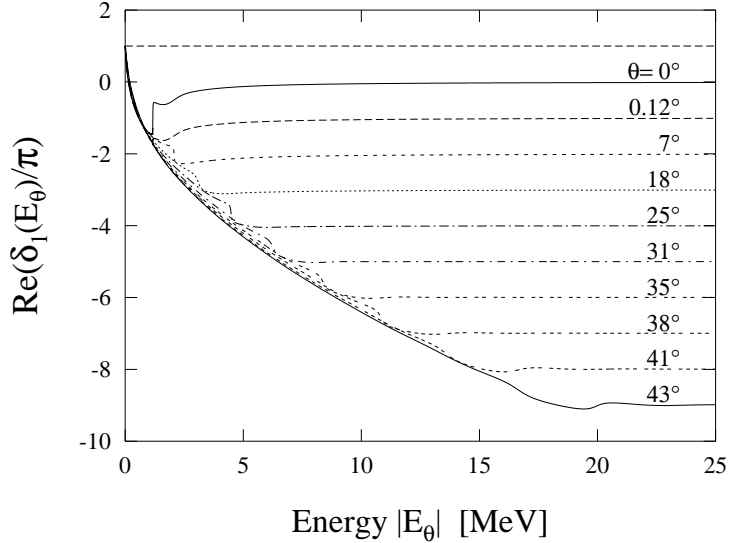}
\caption{Scattering phase shift of $1^-$ states as functions of the energy of rotated continuum states, $E_\theta$ changing the scaling angle $\theta$.}
\label{fig:1-PS}
  \end{center}
\end{figure}

We now discuss the properties of the rotated continuum states in CSM.
For this purpose, we calculate the scattering phase shift of $1^-$ states on the rotated continuum line with energy of 
$E_\theta=E\, e^{-2i\theta}$ by increasing the scaling angle $\theta$.
It is interesting to see the effect of resonances on the phase shift.
For the phase shift $\delta_\ell(E_\theta)$ with the partial wave $\ell$ on the rotated continuum line,
the resonances appearing in a wedge region between the real energy axis and the rotated continuum line 
are treated as bound states as shown in Fig. \ref{fig:csoto-pole}.
From this property, the difference between the real part of $\delta_\ell(E_\theta=0)$ and
that of $\delta_\ell(|E_\theta| \to \infty)$ goes to $\left(N_b + N_r^\theta \right)\pi$,
where $N_b$ and $N_r^\theta $ are the numbers of bound and resonant states obtained with the scaling angle $\theta$, respectively.
Rittby et al. called this property the generalized Levinson theorem \cite{rittby81}.

In Fig.~\ref{fig:1-PS}, we show the phase shifts for $1^-$ states ($\ell=1$) with increasing $\theta$,
in which the real part of the phase shifts is presented. 
We can see one bound state from Fig. \ref{fig:G-pot} (b), $N_b=1$, and the phase shift starts from $\pi$ at $E=0$ and goes to zero at $E=\infty$ when $\theta=0$.
At $E=1.17$ MeV there is a sharp resonance, as shown in Fig. \ref{fig:csoto-pole}, which causes a rapid increase of the phase shift.
This structure disappears when $\theta$ increases and passes beyond $\theta_{\rm c}=0.12^{\circ}$ of its resonance value, because the resonance is decomposed from the rotated continuum states.
Furthermore, a shift by a step of $\pi$ is clearly seen at a large energy for every $\theta$ value, 
which is chosen to be an appropriate angle between resonance positions.
Thus, the asymptotic behavior of $\delta_\ell(E_\theta)$ at large energy,
certainly corresponds to the number of resonances $N_r^\theta$ with the scaling angle $\theta$.

\subsection{Continuum level density}\label{sec:CLD}

The continuum level density (CLD) is an important quantity in the description of quantum scattering phenomena, 
since it is connected with the scattering $S$-matrix  \cite{Le69,tsang75,osborn76}. 
The CLD is defined as the variation of the level densities due to the interaction, and the scattering phase shift ($S$-matrix) is derived from CLD. 
In the analysis of the CLD, we can discuss the relation between resonance solutions and the scattering phase shifts. 
In this section we give an application of CLD in CSM, 
which provides the decomposition of the scattering phase shifts into resonance and non-resonant continuum contributions \cite{odsuren14}.
We also show the application of CLD to a realistic $\alpha$+$n$ system. 

\begin{figure}[b]
\begin{minipage}[b]{0.47\linewidth}
\begin{center}
\includegraphics[width=7.0cm]{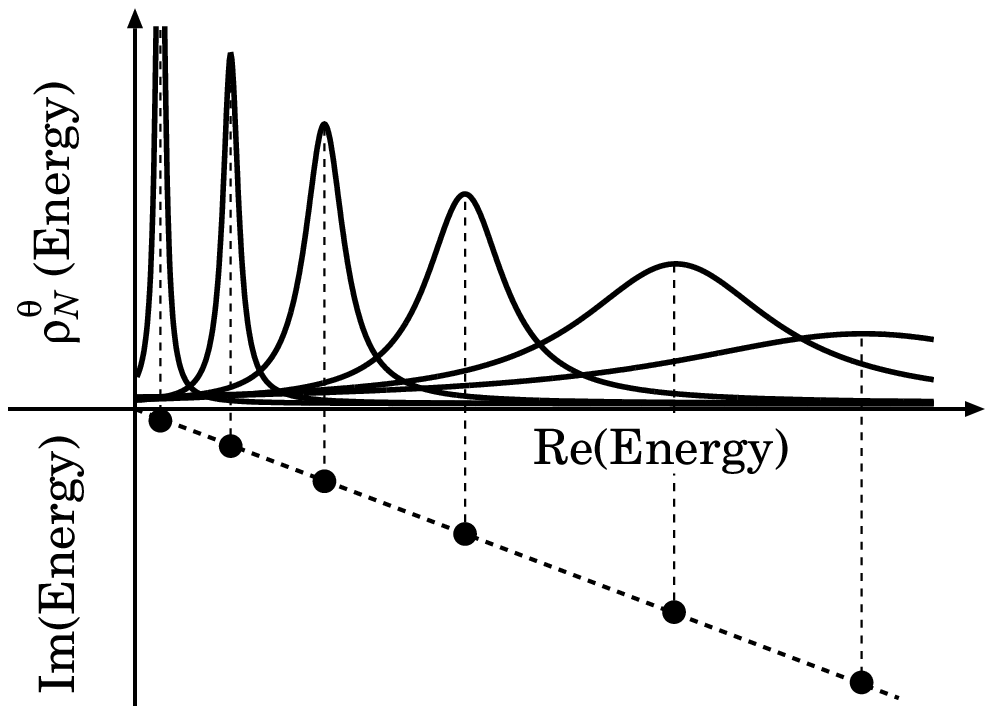}
\caption{Schematic energy eigenvalue distribution with complex-scaled Hamiltonian on a complex energy plane (black circles),
and the contributions of each eigenstate to the level density (solid lines). }  
\label{fig:smooth}
\end{center}
\end{minipage}
\hspace{0.8cm}
\begin{minipage}[b]{0.47\linewidth}
\vspace*{-1cm}
\begin{center}
\includegraphics[width=7.5cm]{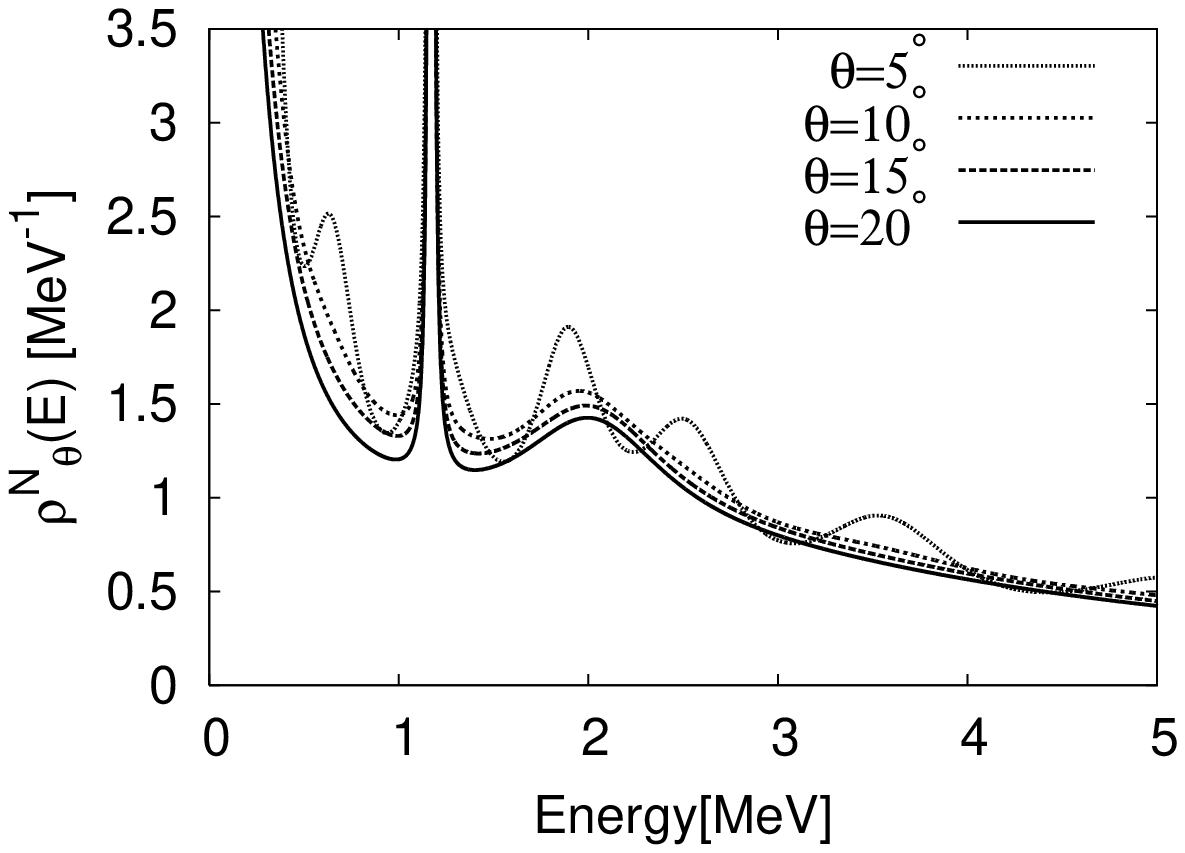}
\caption{Level density $\rho^\theta_N(E)$ with respect to the complex-scaled Hamiltonian $H^\theta$ in the schematic potential model.
Several $\theta$ cases are shown.} 
\label{fig:LD}
\end{center}
\end{minipage}
\end{figure}

In Ref.~\cite{suzuki05} we proposed a method to calculate CLD using CSM in terms of the discretized energy eigenstates.
In this method, the smoothing procedure for each eigenenergy is automatically performed using the scaling angle $\theta$ 
as shown in Fig.~\ref{fig:smooth} without other artificial parameters such as the Strutinsky method \cite{Kr98,strutinsky67}.
The concept of the method is based on the extended completeness relation \cite{My98}, originally proposed by Berggren \cite{Be68}, for bound, resonance, and continuum states in CSM.
The Green's functions can be expressed using ECR in terms of discretized eigenvalues in CSM expanded with a finite number of basis functions.
Because the complex-scaled Hamiltonians $H^\theta$ and $H_0^\theta$ have complex eigenvalues, singularities such as a $\delta$-function are replaced with Lorentzian functions as shown in Fig.~\ref{fig:smooth}.
We show an example of the level density in the schematic potential given in Eq.~(\ref{eq:csoto_pot}).
The level density $\rho^\theta_N(E)$ for the $0^+$ state is shown in Fig.~\ref{fig:LD} using the basis number $N$ by changing $\theta$, where we take $N=30$ to get converging results.
It is found that $\rho^\theta_N(E)$ is smoothed by increasing $\theta$.
It has been shown that the CLD denoted by $\Delta^\theta_N(E)=\rho^\theta_N(E)-\rho^\theta_{0,N}(E)$ is obtained independent of the scaling angle $\theta$ in CSM \cite{suzuki05,suzuki08}.

\begin{figure}[t]
\begin{center}
\includegraphics[width=9.0cm]{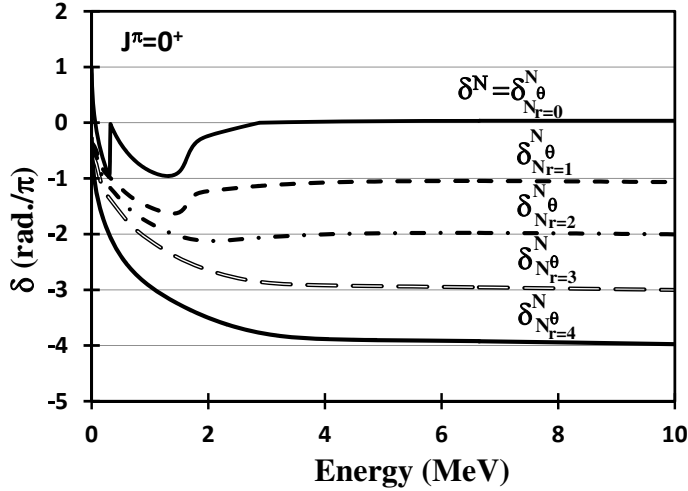}
\caption{Phase shifts of the schematic potential for the $0^{+}$ state and the subtraction of the number of $N_r^\theta$ of resonance terms one by one.}
\label{fig:csoto-PS}
\end{center}
\end{figure}

First, we show the example of the calculation of phase shifts using CLD in the schematic potential.
Using the property of CLD, we calculate the phase shifts by integration from $E=0$ as
\begin{eqnarray}
\delta^N(E)=N_b\pi+\pi\int^{E}_{0}\Delta_N^\theta(E')\, dE'.
\label{eq3-1-6}
\end{eqnarray}
Here $N_b$ is the number of bound states.
In Fig.~\ref{fig:csoto-PS} we show the phase shifts of the $0^+$ state with $N_b=1$. 
In order to see the explicit effect of resonances on the phase shifts, we calculate the phase shifts subtracting the resonance term from $\delta^N(E)$ as
\begin{eqnarray}
\delta^N_{N^\theta_r}(E)&=&\delta^N(E) - \sum_{r=1}^{N^\theta_r}\int^{E}_{0}dE'\frac{\Gamma_r/2}{(E'-E^{\rm res}_r)^2+\Gamma^2_r/4}.
\end{eqnarray}
Here $N^\theta_r$ is the number of resonances, which depends on the scaling angle $\theta$.
The resonance energy and decay width are given as $E^{\rm res}_r$ and $\Gamma_r$, respectively.
In Fig.~\ref{fig:csoto-PS} the results are shown for $N^\theta_r=0,\,1,\,2,\,3,$ and 4.
It is shown that the phase shifts go downward with increasing $N^\theta_r$, in a step of $\pi$ at higher energies from the Levinson theorem.
It is found that the effects of the first and second resonances are remarkable on the structures of the phase shifts with energy up to 2 MeV, but the third and fourth resonances having larger decay widths, do not have notable effects.

Next, we show the realistic example of the unbound $^5$He nucleus as a two-body $\alpha$+$n$ cluster system. 
For the interaction between the $\alpha$ particle and $n$, we use the so-called microscopic KKNN interaction \cite{kanada79}, 
which consists of the central and $LS$ terms and is commonly used in the calculation of He isotopes with the $\alpha$ core plus valence-neutron model in the next section.

\begin{figure}[bh]
\begin{center}
\includegraphics[width=14.0cm]{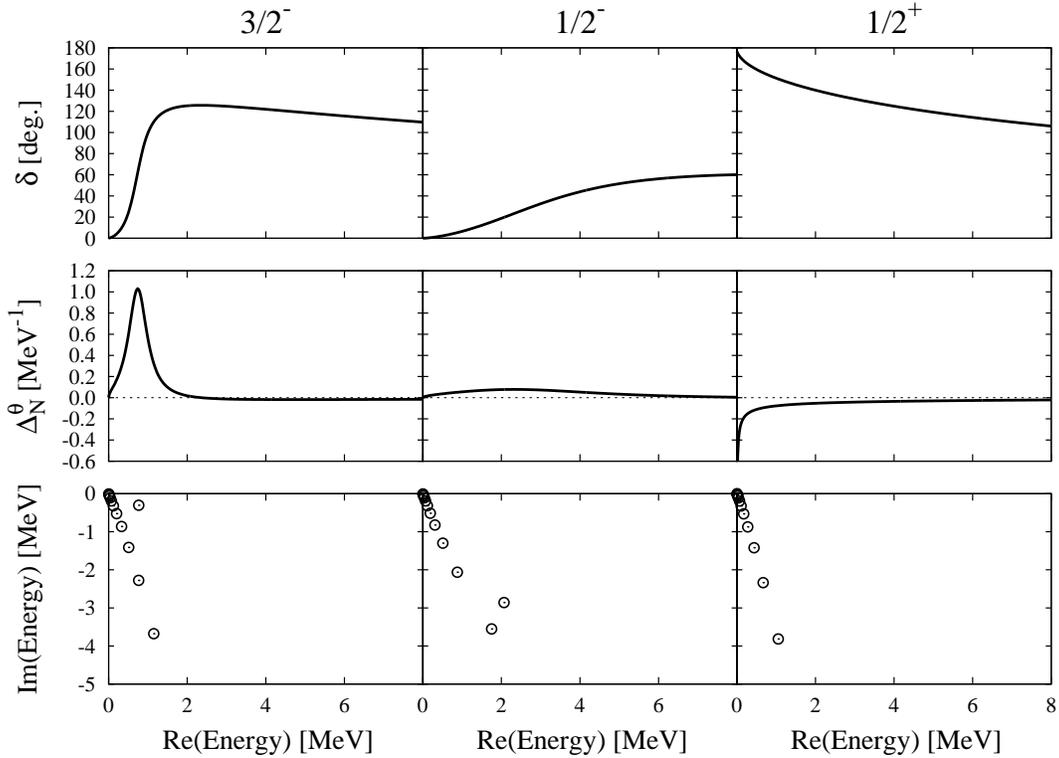}
\caption{Various properties of the $\alpha$+$n$ system with $J^\pi=3/2^-,1/2^-, 1/2^+$ states.
Upper: phase shifts, middle: continuum level density, and lower: energy eigenvalue distributions in a complex energy plane. The scaling angle $\theta$ is 35$^\circ$.}
\label{fig:5He_PS}
\end{center}
\end{figure}

Using the Gaussian basis states, we calculate the energy eigenvalues of the complex-scaled Hamiltonian of $^5$He with $\theta=35^\circ$, and the results for the three states of $3/2^-$, $1/2^-$, and $1/2^+$ are shown in the bottom panels of Fig.~\ref{fig:5He_PS}. 
It is found that the $3/2^-$ and $1/2^-$ states have one resonance pole of the $p$-wave neutron, corresponding to the observed resonances of $^5$He.
One resonance pole of the $\alpha$--$n$ system is obtained, $(E_r^{\rm res},\Gamma_r)=(0.74,~0.59)$ MeV for $p_{3/2}$ and $(2.10,~5.82)$ MeV for $p_{1/2}$, which agree well with the experimental data \cite{aoyama95,Ao06,Ti02}.
The energy difference of two states comes from the $LS$ force in the $\alpha$+$n$ interaction.
The $1/2^+$ state has no resonance of the $s$-wave neutron. 

In addition to resonances, the discretized non-resonant continuum solutions are obtained along the $2\theta$ line as shown in Fig.~\ref{fig:5He_PS}.
Several continuum solutions are off the $2\theta$ line.
This is because that the couplings between resonance and non-resonant continuum states are not accurately described 
due to the finite number of basis functions. 
However, the resonances are solved with appropriate accuracy and the resulting CLD is obtained from these eigenstates satisfactorily, although the positions of some continuum solutions are slightly off the $2\theta$ line.   

\begin{figure}[bh]
\begin{center}
\includegraphics[width=12.0cm]{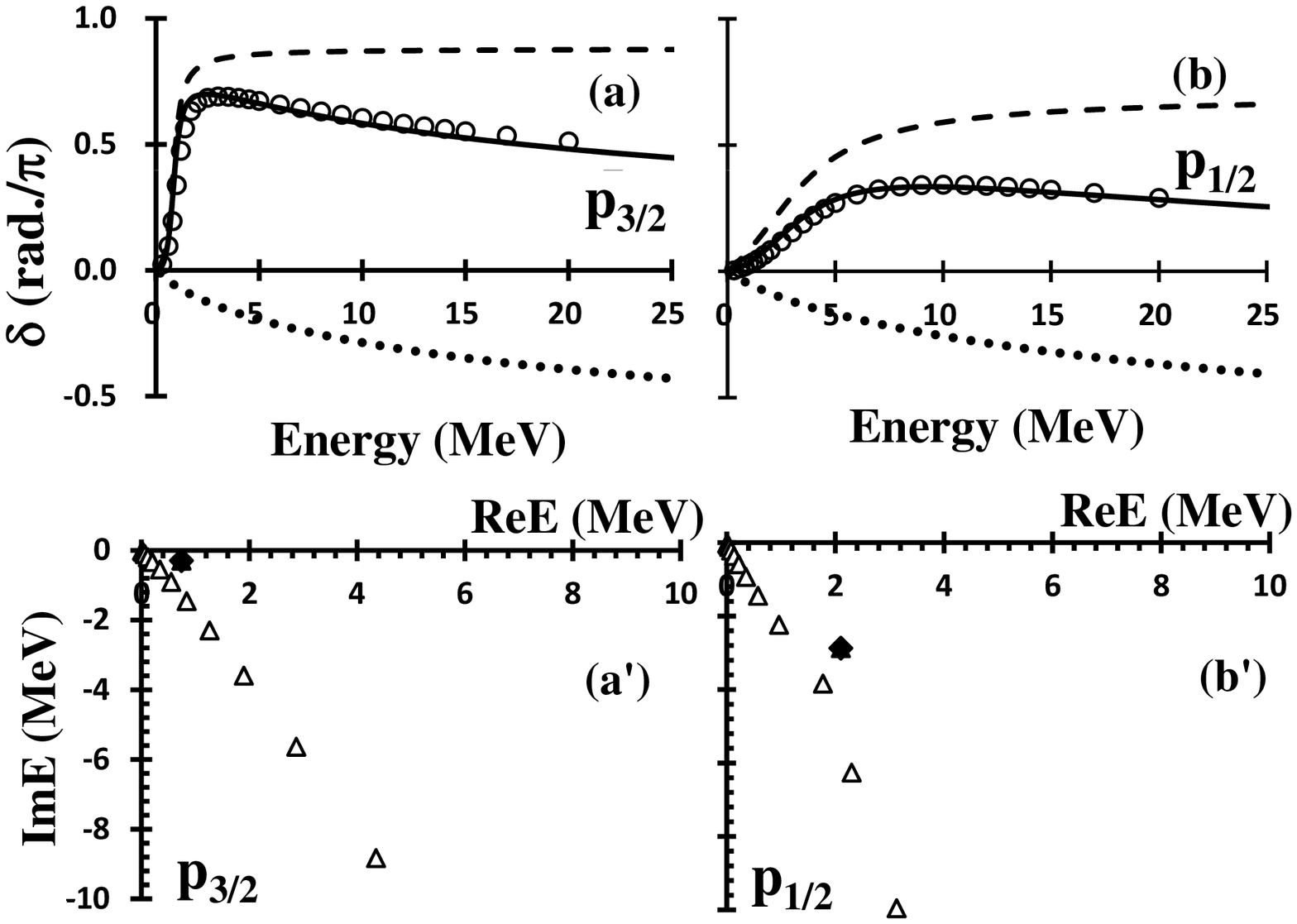}
\caption{Decomposition of the scattering phase shifts of the $\alpha$--$n$ system for (a) $p_{3/2}$ and (b) $p_{1/2}$. The dashed and dotted lines represent the contributions of the resonance and continuum terms, respectively. 
The solid lines display the total scattering phase shifts. 
The experimental data \cite{Ho66} are shown with open circles. }
\label{fig:5He_decompose}
\end{center}
\end{figure}

According to ECR for $\alpha$+$n$ system, we can decompose CLD and the phase shifts into resonance and non-resonant components,
as shown in Fig.~\ref{fig:5He_decompose} together with experimental data. We can see good agreement between the theoretical and experimental results. 
The resonance component of the phase shift of $p_{3/2}$ increases rapidly due to the small decay width. 
Although $p_{1/2}$ has a larger width, the phase shift of $p_{1/2}$ shows clear resonance behavior. 
The continuum components (dotted lines) of phase shifts for both states are very similar. 
This behavior seems due to the same $p$-wave scattering and a small effect of the $LS$ force on the background states.
The present $\alpha$--$n$ systems show clear resonance behavior. 
We also investigate similar analysis for the $\alpha$--$\alpha$ two-cluster system \cite{odsuren14}.
The analysis of CLD in the three-body system is straightforward, as for the triple-$\alpha$ states in $^{12}$C \cite{My14,kurokawa05}.

\clearpage                                                                                
\section{Many-body resonances and non-resonant continuum states} \label{sec:many}
We apply CSM to the resonances phenomena observed in unstable nuclei, 
in particular many-body resonances beyond the two-body case, which are located above the threshold energy of the many-body decay.
We take the case of the structures of neutron-rich He isotopes and their mirror proton-rich nuclei, most states of which 
are unbound due to the weak binding property of valence nucleons to the $\alpha$ particle.
In the present nuclear model, we treat the $\alpha$ particle as a frozen core nucleus
 and solve the motions of valence neutrons/protons surrounding the $\alpha$ core for neutron-rich He isotopes/proton-rich cases.

\subsection{Cluster orbital shell model} \label{sec:COSM}
\begin{figure}[b]
\centering
\includegraphics[width=14.0cm,clip]{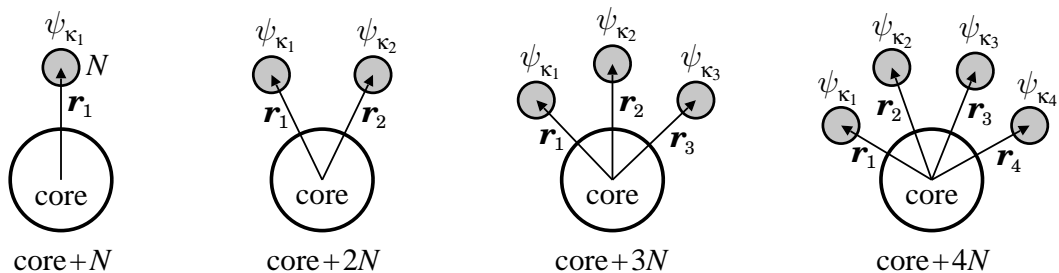}
\caption{Set of spatial coordinates in COSM consisting of a core nucleus and valence nucleons $N$.}
\label{fig:COSM}
\end{figure}

We explain the nuclear model describing He isotopes and their mirror nuclei on the basis of the cluster model assuming an $\alpha$ core.
We use the so-called cluster orbital shell model (COSM) \cite{suzuki88}.
In COSM, the motion of valence nucleons around a core nucleus is solved as shown in Fig. \ref{fig:COSM}, which has an analogy with the single-particle motion of the shell model.
In addition, COSM has the merit of the cluster model to obtain the relative wave function of valence nucleons with respect to the core nucleus precisely.
This COSM is extendable to treat excitations of the core nucleus in terms of a multi-configuration representation \cite{myo07b,myo08}.

The Hamiltonian for COSM consisting of the $\alpha$ particle with a mass number $A_{\rm c}=4$ and $N_{\rm v}$ valence nucleons \cite{masui06,myo07a,myo09b,myo10} is
\begin{eqnarray}
	H
&=&	\sum_{i=0}^{N_{\rm v}} t_i - T_{\rm c.m.} + \sum_{i=1}^{N_{\rm v}} V^{\alpha N}_i + \sum_{i<j}^{N_{\rm v}}   V^{NN}_{ij} 	\\
&=&	\sum_{i=1}^{N_{\rm v}} \left[ \frac{\vc{p}^{ 2}_i}{2\mu} + V^{\alpha N}_i \right] + \sum_{i<j}^{N_{\rm v}} \left[ \frac{\vc{p}_i\cdot \vc{p}_j}{A_{\rm c} m} + V^{NN}_{ij} \right] ,
        \label{eq:COSM_ham}
\end{eqnarray}
where $t_i$ and $T_{\rm c.m.}$ are the kinetic energies of each particle ($\alpha$ and nucleon $N$) and the center of mass of the total system with a mass number of $A_{\rm c}+N_{\rm v}$, respectively.
The term $t_0$ is for the $\alpha$ core nucleus.
The operator $\vc{p}_i$ is the relative momentum between $\alpha$ and a valence nucleon. 
The reduced mass $\mu$ is $A_{\rm c} m/(A_{\rm c}+1)$ with a nucleon mass $m$. 
The potentials $V^{\alpha N}$ and $V^{NN}$ represent the interactions between $\alpha$ particle and a nucleon, and between valence nucleons.

In COSM, the total wave function $\Psi^{JT}_{\rm COSM}$ with total mass number $A=A_{\rm c}+N_{\rm v}$, spin $J$, and isospin $T$ is represented by the superposition of the different configurations $\Phi^{JT}_c$ as
\begin{eqnarray}
    \Psi_{\rm COSM}^{JT}
= \sum_c C^{JT}_c\, \Phi^{JT}_c(A),
    \label{eq:COSM-WF0}
    \qquad
    \Phi^{JT}_c(A)
= \prod_{i=1}^{N_{\rm v}} a^\dagger_{\kappa_i}|0\rangle , 
    \label{eq:COSM-WF1}
\end{eqnarray}
where the vacuum $|0\rangle$ represents the $\alpha$ particle with spin $J=0$ and isospin $T=0$.
The creation operator $a^\dagger_{\kappa}$ is for the single-particle state of a valence nucleon above the $\alpha$ core with the $(0s)^4$ closed configuration of a harmonic oscillator wave function.
The quantum number $\kappa$ is a set of $\{n,\ell,j,t_z\}$, where
the index $n$ represents the different radial component and $\ell$ is the orbital angular momentum of a valence nucleon.
The $z$ component of the isospin of each nucleon (proton or neutron) is given as $t_z$. 
The index $c$ represents the set of $\kappa_i$ as $c=\{\kappa_1,\ldots,\kappa_{N_{\rm v}}\}$ for all valence nucleons,
which determines the configuration of the COSM wave function.
The expansion coefficients $\{C_c^{JT}\}$ in Eq.~(\ref{eq:COSM-WF0}) are determined 
by diagonalization of the Hamiltonian matrix with complex scaling.

The coordinate representation of the single-particle state $a^\dagger_{\kappa}$ is given as 
$\psi_{\kappa}(\vc{r})$ as functions of the relative coordinate $\vc{r}$ between the center of mass of the $\alpha$ and a valence nucleon \cite{suzuki88}.
An illustration of the coordinate set is shown in Fig.~\ref{fig:COSM} up to the $N_{\rm v}=4$ case.
In COSM, we expand the radial part of $\psi_\kappa(\vc{r})$ using the Gaussian basis functions for each orbit \cite{Ao06,Hi03} as
\begin{eqnarray}
    \psi_\kappa(\vc{r})
&=& \sum_{k=1}^{N_{\ell j}} d^k_{\kappa}\ \phi_{\ell j t_z}^k(\vc{r},b_{\ell j}^k),
    \label{eq:COSM-base1}
    \\
    \phi_{\ell j t_z}^k(\vc{r},b_{\ell j}^k)
&=& {\cal N}\, r^{\ell} e^{-(r/b_{\ell j}^k)^2/2}\, [Y_{\ell}(\hat{\vc{r}}),\chi^\sigma_{1/2}]_{j}\, \chi^\tau_{t_z},
    \label{eq:Gauss}
	\\
    \langle \psi_\kappa|\psi_{\kappa'} \rangle 
&=& \delta_{\kappa,\kappa'},
    \label{eq:COSM-base2}
\end{eqnarray}
where the index $k$ is to distinguish the range parameter $b_{\ell j}^k$ of the Gaussian functions with the number $N_{\ell j}$, which is determined to converge the physical solutions.
The length parameters $b_{\ell j}^n$ are typically chosen in a geometric progression \cite{Ao06,Ka88,Ka89,myo09b}.
The normalization factors of the basis are given by ${\cal N}$.
The coefficients $\{d^k_{\kappa}\}$ in Eq.~(\ref{eq:COSM-base1}) are determined in the Gram--Schmidt orthonormalization,
and the basis states $\psi_\kappa$ are orthogonal to each other in Eq.~(\ref{eq:COSM-base2}).
The same technique using Gaussian bases as a single--particle basis is used in the tensor-optimized shell model \cite{myo07b,myo05,myo09a,myo11a,myo12a} and tensor-optimized antisymmetrized molecular dynamics \cite{myo15,myo17a} for the analysis of light nuclei with bare nuclear forces.

The antisymmetrization between the $\alpha$ core and a valence nucleon is described in the orthogonality condition model \cite{Ao06}, 
where the single particle states $\psi_{\kappa}$ are orthogonal to the $0s$ state in the $\alpha$ core.

We apply CSM to the many-body COSM wave function.
In CSM, all of the relative coordinates $\vc{r}_i$ between $\alpha$ and a valence nucleon are complex-scaled into $\vc{r}_i\, e^{i\theta}$
for $i=1,\ldots,N_{\rm v}$ with a common scaling angle $\theta$.
The Hamiltonian in Eq.~(\ref{eq:COSM_ham}) becomes the complex-scaled Hamiltonian $H_{\rm COSM}^\theta$, and the complex-scaled Schr\"odinger equation is written as
\begin{eqnarray}
    H^\theta\, \Psi^{JT,\, \theta}_{\rm COSM}
&=& E_{JT}^\theta\, \Psi^{JT,\, \theta}_{\rm COSM} ,
	\\
    \Psi_{\rm COSM}^{JT,\,\theta}
&=& \sum_c C^{JT,\, \theta}_c \Phi^{JT}_c(A),
	\\
H^\theta 
&=&  U(\theta)\,H\, U^{-1}(\theta).
\end{eqnarray}
The expansion coefficients $C^{JT,\, \theta}_c$ depend on $\theta$ and are obtained from the eigenvalue problem of $H^\theta$ with the COSM basis functions
using the complex-scaled Hamiltonian matrix elements;
\begin{eqnarray}
 \langle \widetilde \Phi^{JT}_c(A) | H^\theta | \Phi^{JT}_{c'}(A) \rangle
 &=&
 \langle \widetilde \Phi^{JT}_c(A)^{-\theta} | H | \Phi^{JT}_{c'}(A)^{-\theta} \rangle ,
\end{eqnarray}
where the biorthogonal relation is taken in the wave functions as explained in Eq.~(\ref{eq-2-1-10}).
The inverse scaling of the basis states with $-\theta$ is treated in the Gaussian basis functions transforming 
$r/b^k_{\ell j}$ to $r/(b^k_{\ell j}e^{i\theta})$ in Eq~(\ref{eq:Gauss}).

The energy eigenvalues $E^\theta_{JT}$ are obtained on a complex energy plane for each spin $J$ and isospin $T$.
We employ a finite number of basis states, which bring the discretized representation of the continuum states as well as the resonances in CSM.

After solving the eigenvalue problem with COSM, we can categorize each eigenstate $\Psi_\nu$ 
into bound, resonance and non-resonant continuum states, which constructs the extended completeness relation (ECR).
We explain the case of $^6$He with isospin $T=1$ consisting of the $\alpha$ core and two valence neutrons.
The various states of $^6$He consist of the three-body ECR \cite{Be68,myo01} as
\begin{eqnarray}
	{\bf 1}
&=&	\sum_{~\nu} \kets{\Psi_\nu}\bras{\wtil{\Psi}_\nu}
        \\
&=&	\{\mbox{Three-body bound states of $^6$He}\}
        \nonumber\\
&+&	\{\mbox{Three-body resonances of $^6$He}\}
        \nonumber\\
&+&	\{\mbox{Two-body continuum states consisting of $^5$He$^{(*)}$+$n$}\}
        \nonumber\\
&+&	\{\mbox{Three-body continuum states consisting of $\alpha$+$n$+$n$}\}
        \label{eq:ECR3}
\end{eqnarray}
where $\{ \Psi_\nu,\wtil{\Psi}_\nu \}$ consist of a set of biorthogonal bases with a state $\nu$.
Note that the $^5$He nucleus, a subsystem of $^6$He, has resonances but no bound states.
The above relation is used to calculate the CSGF in Eq. (\ref{eq-2-3-1}) and the transition strength into the three-body unbound states of $^6$He
such as the $E1$ and $E2$ transitions \cite{myo01} and also the breakup reactions \cite{matsumoto10}.

\subsection{He isotopes and their mirror nuclei} \label{sec:He-COSM}

We discuss the spectroscopy of He isotopes and their mirror nuclei with COSM.
In the Hamiltonian of COSM, two kinds of the interactions between core--$N$ and $N$--$N$ are necessary.
In the present study, the $\alpha$-$n$ interaction $V^{\alpha n}$ is given by the microscopic KKNN potential \cite{aoyama95,kanada79}, also used in the previous section.
We use the effective Minnesota central potential \cite{tang78} as the nuclear part of $V^{NN}$ in addition to the Coulomb interaction.

We explain the model space of COSM for He isotopes \cite{myo07b,myo09b,myo10}.
For the single-particle states, we take the angular momenta $\ell\le 2$ and adjust $V^{NN}$ slightly to reproduce the two-neutron separation energy of $^6$He ($0^+$) from experiment of 0.975 MeV, which is small for the neutron-halo structure in the ground state.

\begin{figure}[t]
\centering
\includegraphics[width=7.4cm,clip]{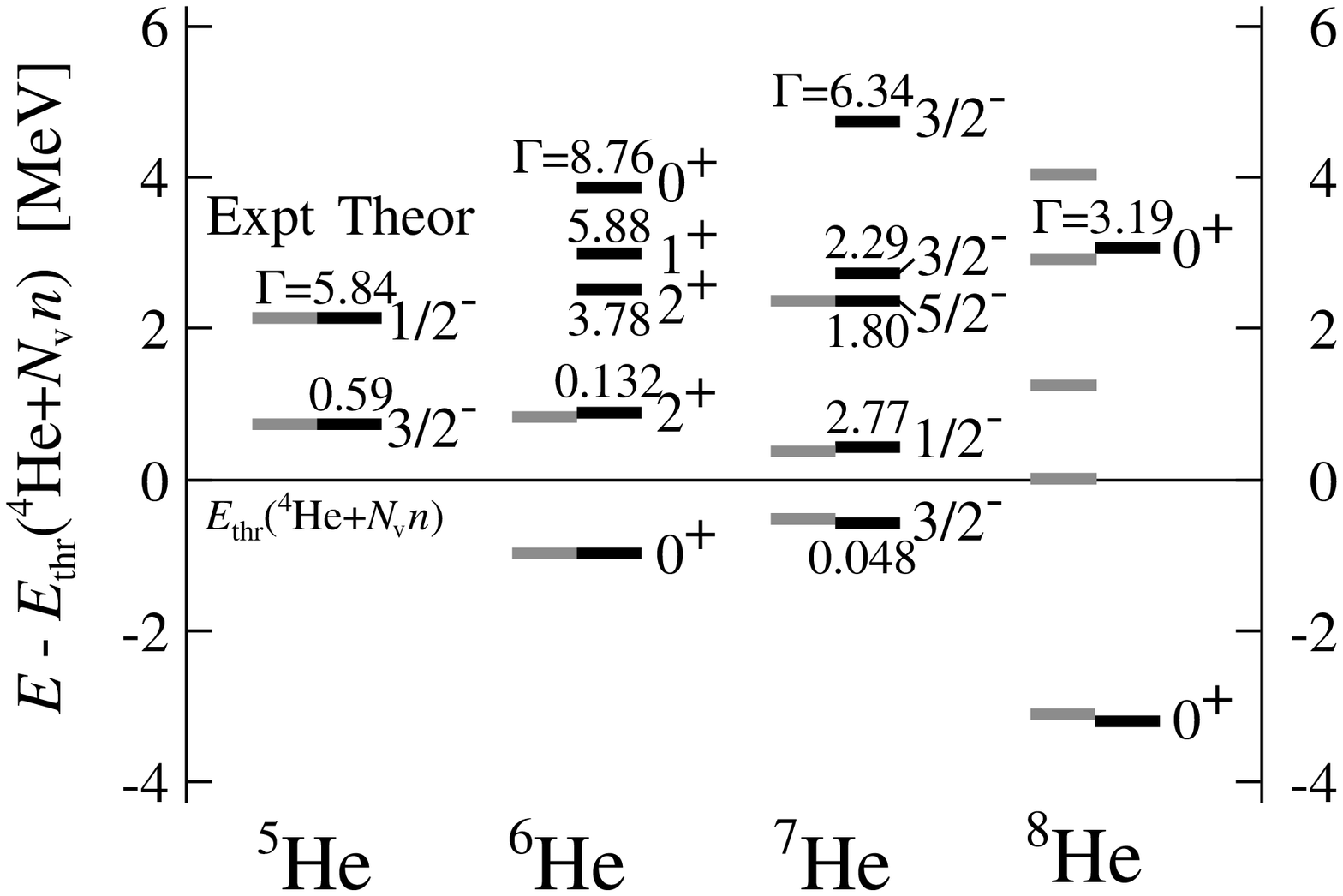}\hspace*{0.4cm}
\includegraphics[width=7.4cm,clip]{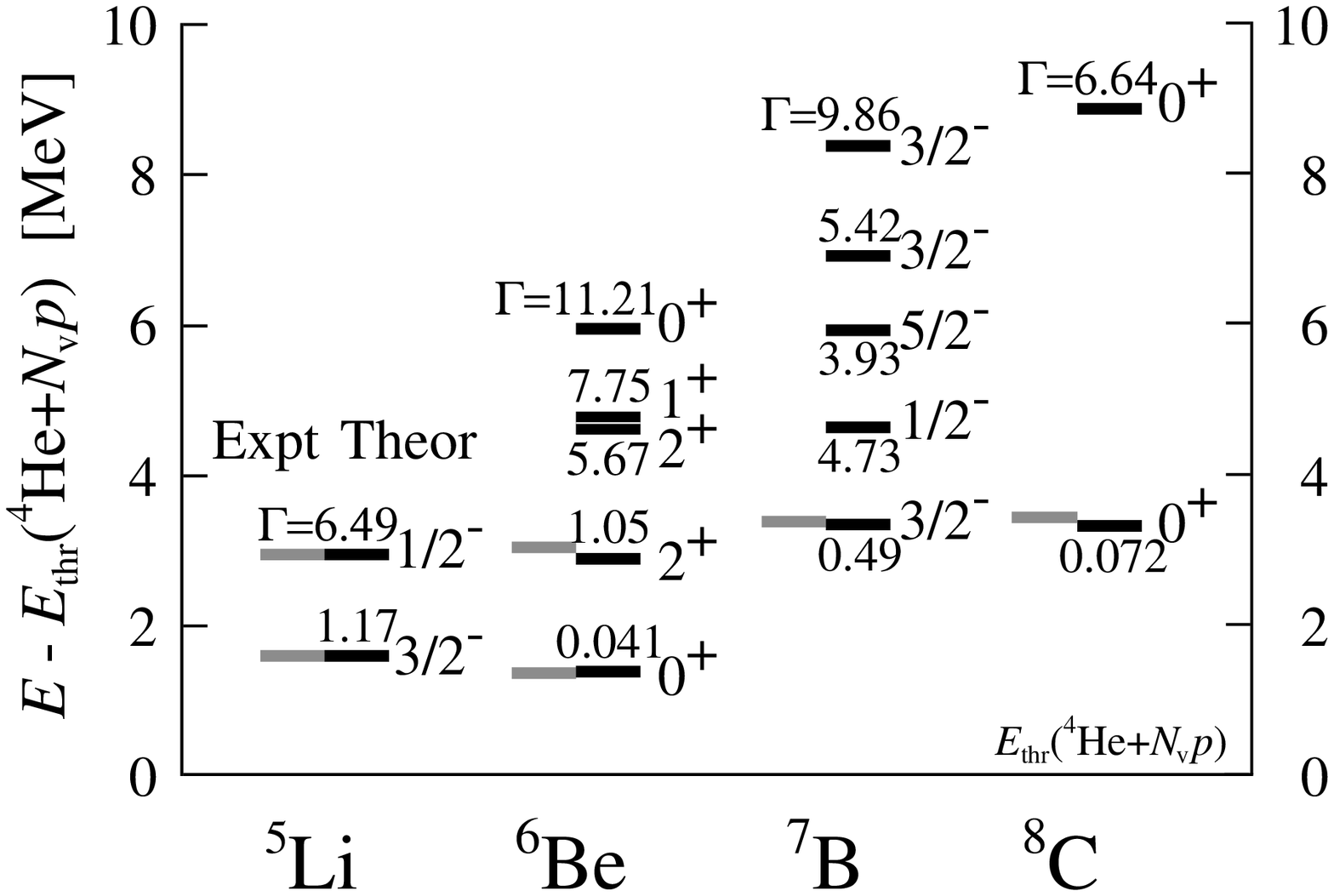}
\caption{
Energy levels of He isotopes (left) and their mirror nuclei (right) measured from the threshold energy of the $\alpha$ particle emission in units of MeV.
The black and gray lines indicate the theoretical and experimental values, respectively. The small numbers are the decay widths in theory. 
The $0^+$ states are shown in theory for $^8$He and $^8$C.}
\label{fig:ene_COSM}
\end{figure}

\begin{figure}[t]
\centering
\includegraphics[width=9.5cm,clip]{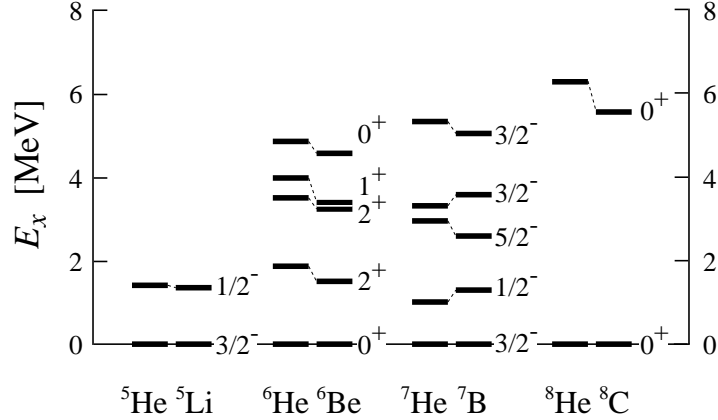}
\caption{Excitation energy spectra of mirror nuclei of $A=5,6,7$, and 8 in units of MeV.}
\label{fig:excite_COSM}
\end{figure}

We show the energy levels of He isotopes and their mirror nuclei with COSM in Fig.~\ref{fig:ene_COSM}, measured from the energy of the $\alpha$ particle \cite{myo117,myo128,myo13}.
The small numbers near the levels represent the decay widths of the states.
One can see good agreement for the energy position between theory and experiment up to the five-body case of $^8$He and $^8$C with isospin $T=2$, 
in which only the $0^+$ states are shown.
It is found that the order of energy levels are the same between proton-rich and neutron-rich sides.
Theory also gives many predictions of the excited states.
In Fig. \ref{fig:excite_COSM}, we compare the excitation energy spectra of proton-rich and neutron-rich sides
to examine the mirror symmetry in nuclei.
It is found that the good symmetry is confirmed between the corresponding nuclei.
The differences in excitation energies for individual levels are less than 1 MeV \cite{myo128}.

The matter and charge radii of the ground states of $^6$He and $^8$He provide good information on the valence neutron distribution.
These values obtained in COSM reproduce the recent experiments, as shown in Table~\ref{tab:sec3_radius_COSM}.
From the results, the COSM wave functions can explain the special extensions of neutrons in the halo and skin structures of He isotopes. 


\nc{\lw}[1]{\smash{\lower1.5ex\hbox{#1}}}
\begin{table}[thb]
\caption{Matter ($R_{\rm m}$) and charge ($R_{\rm ch}$) radii of $^6$He and $^8$He obtained with COSM in units of fm.
Experimental values are taken from a\cite{tanihata92}, b\cite{alkazov97}, c\cite{kiselev05}, d\cite{mueller07}, and e\cite{brodeur12}.}
\label{tab:sec3_radius_COSM}
\centering
\begin{tabular}{r p{1.8cm} p{5.0cm}}
\hline\hline
                           & COSM   & Experiments        \\ 
\hline
\lw{$^6$He}~~~$R_{\rm m}$  &~~2.37  & 2.33(4)$^{\rm a}$~~~~2.30(7)$^{\rm b}$~~~~2.37(5)$^{\rm c}$ \\
              $R_{\rm ch}$ &~~2.01  & 2.068(11)$^{\rm d}$ \\
\hline
\lw{$^8$He}~~~$R_{\rm m}$  &~~2.52  & 2.49(4)$^{\rm a}$~~~~2.53(8)$^{\rm b}$~~~~2.49(4)$^{\rm c}$ \\
              $R_{\rm ch}$ &~~1.92  & 1.929(26)$^{\rm d}$~~~~1.959(16)$^{\rm e}$ \\
\hline
\end{tabular}
\end{table}

\subsection{$2^+$ resonances and continuum states in $^6$He}
\begin{figure}[bht]
\centering
\includegraphics[width=7.5cm,clip]{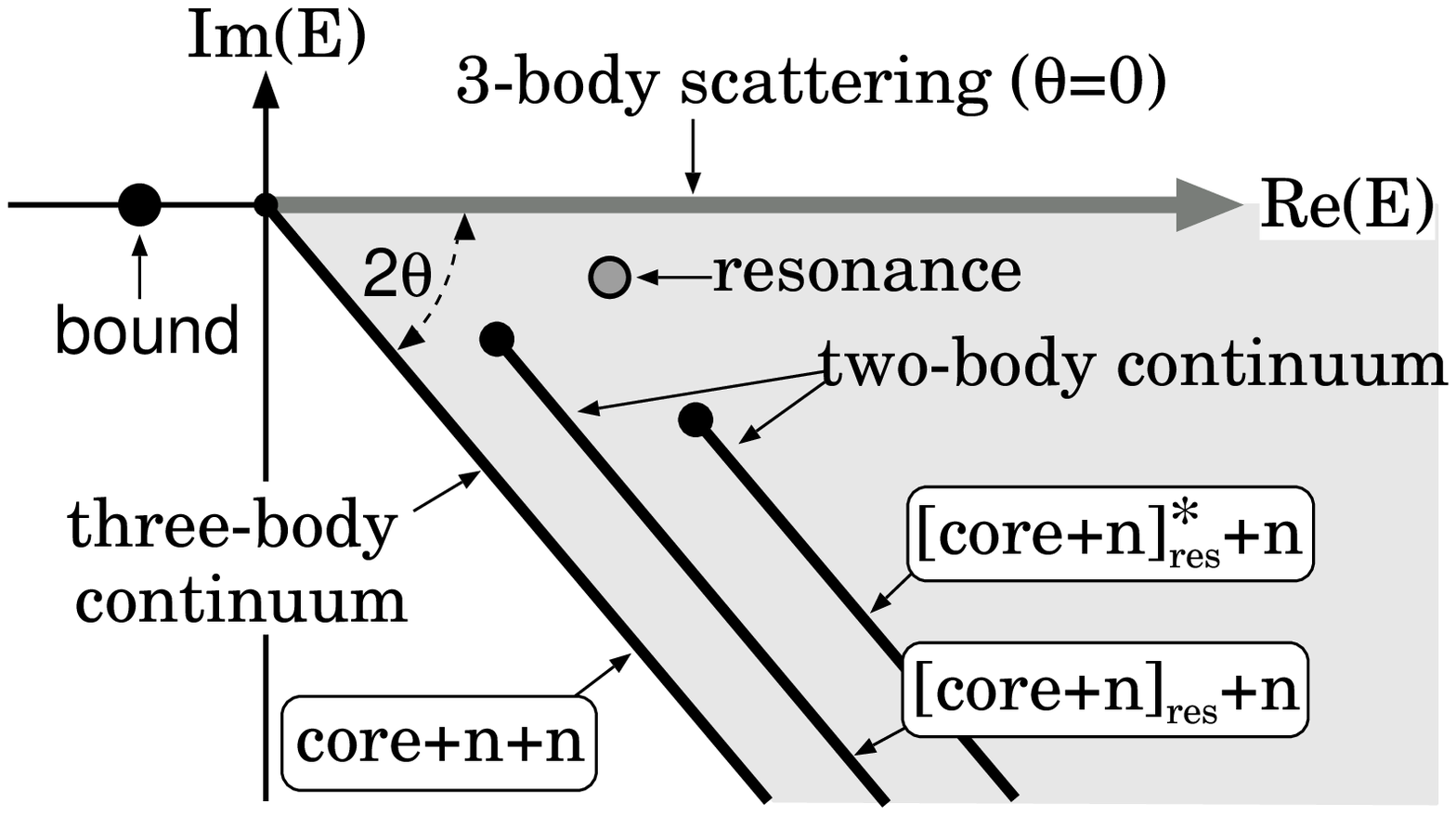}\hspace*{0.5cm}
\includegraphics[width=8.0cm,clip]{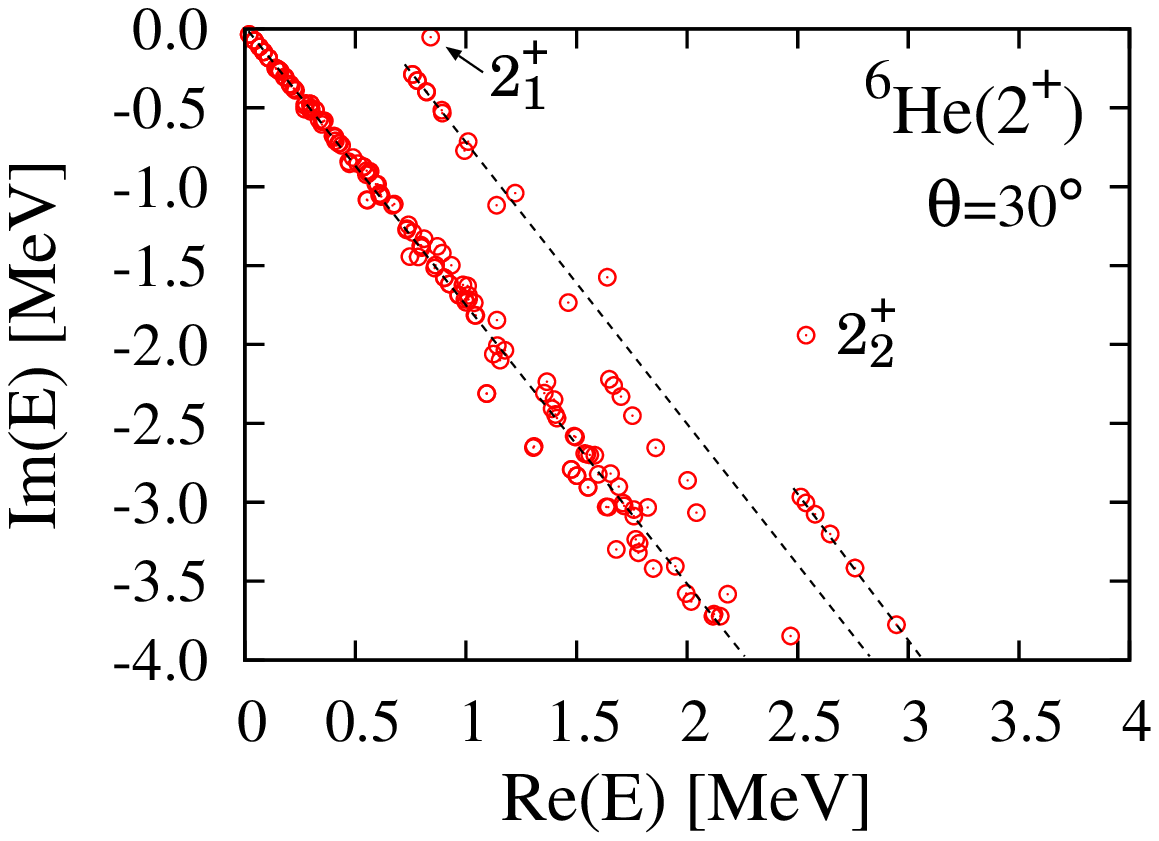}
\caption{
Left: Energy eigenvalue distribution of core+$n$+$n$ in CSM.
Right: Energy eigenvalues of the $^6$He($2^+$) states with $\theta=30^\circ$ in the complex energy plane measured from the $\alpha$+$n$+$n$ threshold energy \cite{myo09b}. 
The three schematic lines with dots indicate the continuum states of $\alpha$+$n$+$n$ and $^5$He($3/2^-$,$1/2^-$)+$n$, from left to right, respectively.}
\label{fig:ene6_2}
\end{figure}

For resonance and non-resonant continuum states in CSM,
the energy eigenvalue distribution of $^6$He ($2^+$) is shown in Fig.\ref{fig:ene6_2}.
In this figure, we can identify the locations of various kinds of unbound state: 
the three-body resonances of $^6$He ($2^+_1$ and $2^+_2$), and
two kinds of continuum states of $^5$He($3/2^-$,$1/2^-$)+$n$ and $\alpha$+$n$+$n$.

\begin{figure}[bht]
\centering
\includegraphics[width=8.5cm,clip]{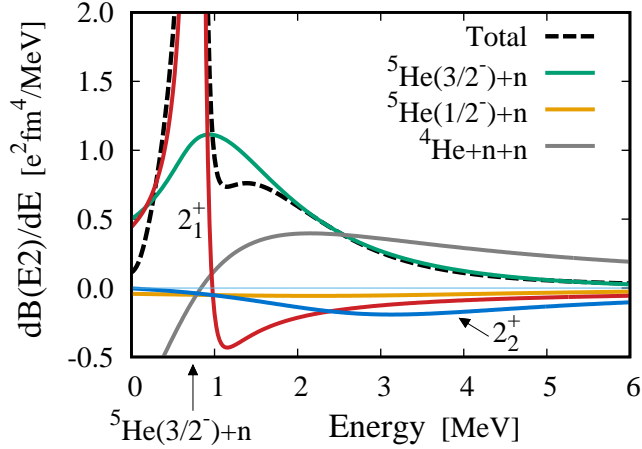}
\caption{$E2$ transition of $^6$He from the ground state to the $2^+$ scattering state in CSM, measured from the $\alpha$+$n$+$n$ threshold energy.}
\label{fig:6He_E2}
\end{figure}

Corresponding to the $2^+$ energies in Fig.~\ref{fig:ene6_2}, we calculate the $E2$ transition strength of $^6$He
excited from the $0^+$ ground state to the three-body scattering states of $\alpha$+$n$+$n$.
It is interesting to investigate the contributions of resonances and non-resonant continuum states.
Before showing the results of the $E2$ transition strength distribution, 
we discuss the transition matrix elements between the ground state and the $2^+_{1,2}$ resonances,
the squared values of which are $2.78+i\;0.43$ for $2^+_1$ and $-1.5+i\;0.9$ for $2^+_2$ in units of e$^2$fm$^4$.
Transition strengths for resonances are obtained as complex numbers as well as the energy eigenvalues. 
When the resonance pole is very close to the real energy axis as for $2^+_1$ with $\Gamma=0.132$ MeV, 
its matrix elements show large real parts.
On the other hand, the $2^+_2$ resonance has a large decay width of 3.78 MeV and a large imaginary part
of the transition strength.

The obtained $E2$ transition strength distribution including resonances and continuum states 
is shown in Fig.~\ref{fig:6He_E2}.
There are five kinds of components: transitions to $2^+_{1,2}$ resonances,
two-body continuum states of $^5$He($3/2^-$,$1/2^-$)+$n$, and three-body continuum ones of $\alpha$+$n$+$n$.
From Fig.~\ref{fig:6He_E2}, we can see that the $2^+_1$ resonance gives the main contribution showing
a sharp peak around the resonance energy of 0.81 MeV with a small decay width. 
On the other hand, the contribution from the $2^+_2$ resonance is very small due to the large decay width.
The components associated with two- and three-body continua are smaller than those of $2^+_1$. 
However, in the continuum transitions the two-body continuum component of $^5$He($3/2^-$)+$n$ shows 
a peak at around 1 MeV, just above the two-body threshold energy of this channel. 
This component mainly contributes to making shoulder structure in the total strength at around 1.5 MeV 
measured from the $\alpha$+$n$+$n$ threshold energy.

The reason why the two-body component of $^5$He($3/2^-$)+$n$ makes a peak at around 1 MeV is as follows:
In this two-body component, the last neutron is in the continuum state and 
we have confirmed that this neutron has a similar structure to the plane wave \cite{myo01}. 
Hence the peak at around 1 MeV is interpreted as a two-body threshold effect 
reflecting the spatially extended structure of neutrons in the initial neutron-halo state. 
A similar peak structure can be observed in the case of $E1$ transitions into $1^-$ states,
which are dominated by the $^5$He($3/2^-$)+$n$ channel and no three-body resonances.
The effect of two-body continuum components of $^5$He(1/2$^-$)+$n$ is very small.
It is found that the three-body continuum component of $\alpha$+$n$+$n$ 
also contributes to making a shoulder at around 1.5 MeV in the total transition strength.

In the calculation, it is found that the $E2$ transition strength distribution 
shows clear resonance behavior of the $2^+_1$ state, but not of the $2^+_2$ state.
It might be hard to observe experimentally the existence of the $2^+_2$ state from the $E2$ transition. 
On the other hand, the continuum strengths consisting of $^5$He(3/2$^-$)+$n$ 
and $\alpha$+$n$+$n$ channels contribute to making a shoulder-like structure. 

\subsection{Mirror symmetry breaking in He isotopes and their mirror nuclei}

We compare the structures of proton- and neutron-rich nuclei from the viewpoint of mirror symmetry
and focus on two nuclei, $^8$He and $^8$C, having four neutrons/protons above the $\alpha$ particle, respectively.
In the two nuclei, only the ground state of $^8$He is a bound state as seen in Fig.\ref{fig:ene_COSM}.
It is interesting to investigate the effect of the Coulomb interaction on the mirror symmetry of these nuclei.
In general, the structures of resonances and weakly bound states are influenced by the open channels of the particle emissions.
In unstable nuclei, the mirror symmetry is considered to be closely related to the coupling between the open channels and the continuum states.
We have also performed a similar analysis on the comparison of $^6$He and $^6$Be with two valence nucleons \cite{myo14}.

We first discuss the configuration mixing of the single-particle states in $^8$He and $^8$C,
the results of which are shown in Table~\ref{comp8_1} for the ground states of $^8$He and $^8$C \cite{myo128}.
We show the dominant configurations and the corresponding squared amplitudes $(C^{JT}_c)^2$ defined in Eq. (\ref{eq:COSM-WF0}).   
The states of $^8$He and $^8$C show similar values for mixing, and the $(p_{3/2})^4$ configuration dominates their wave functions.

We can see the configuration mixing of the $0^+_2$ resonances of $^8$He and $^8$C in Table \ref{comp8_2} \cite{myo128}.
It is found that the $(p_{3/2})^2(p_{1/2})^2$ configuration dominates the wave functions of $^8$He and $^8$C with large squared amplitudes of about 0.93--0.97 in the real parts.
These states can be regarded as the two-particle two-hole excitations from the ground states.
The magnitudes of the dominant configuration in $0^+_2$ are larger than those of the ground states of $^8$He and $^8$C commonly.
This mainly comes from the reduction of the coupling strengths between different configurations for valence nucleons \cite{myo14}.
When the $0^+_2$ resonances of $^8$He and $^8$C are spatially extended,
their wave functions are widely distributed and the amplitudes of valence nucleons easily penetrate from the interaction range to the outside.
From this effect, the couplings between the different configurations are reduced and the single configuration of $(p_{3/2})^2(p_{1/2})^2$ largely survives in the $0^+_2$ states.

\begin{table}[t]
\caption{Squared amplitudes $(C^{JT}_c)^2$ of the ground states of $^8$He and $^8$C.}
\label{comp8_1}
\centering
\begin{tabular}{c|ccc}
\noalign{\hrule height 0.5pt}
Configuration              & $^8$He ($0^+_1$)& $^8$C ($0^+_1$)   \\
\noalign{\hrule height 0.5pt}
 $(p_{3/2})^4$             & 0.860           & $0.878-0.005{\it i}$  \\
 $(p_{3/2})^2(p_{1/2})^2$  & 0.069           & $0.057+0.001{\it i}$  \\
 $(p_{3/2})^2(1s_{1/2})^2$ & 0.006           & $0.010+0.003{\it i}$  \\
 $(p_{3/2})^2(d_{3/2})^2$  & 0.008           & $0.007+0.000{\it i}$  \\
 $(p_{3/2})^2(d_{5/2})^2$  & 0.042           & $0.037+0.000{\it i}$  \\
\noalign{\hrule height 0.5pt}
\end{tabular}
\end{table}

\begin{table}[t]
\caption{Squared amplitudes $(C^{JT}_c)^2$ of the $0^+_2$ states of $^8$He and $^8$C.}
\label{comp8_2}
\centering
\begin{tabular}{c|ccc}
\noalign{\hrule height 0.5pt}
Configuration             &  $^8$He ($0^+_2$)      & $^8$C ($0^+_2$)  \\
\noalign{\hrule height 0.5pt}
$(p_{3/2})^4$             &  $0.020-0.009{\it i}$  & $ 0.044+0.007{\it i}$ \\
$(p_{3/2})^2(p_{1/2})^2$  &  $0.969-0.011{\it i}$  & $ 0.934-0.012{\it i}$ \\
$(p_{3/2})^2(1s_{1/2})^2$ & $-0.010-0.001{\it i}$  & $-0.001+0.000{\it i}$ \\
$(p_{3/2})^2(d_{3/2})^2$  &  $0.018+0.022{\it i}$  & $ 0.020+0.003{\it i}$ \\
$(p_{3/2})^2(d_{5/2})^2$  &  $0.002+0.000{\it i}$  & $ 0.002+0.001{\it i}$ \\
\noalign{\hrule height 0.5pt}
\end{tabular}
\end{table}

We next discuss the effect of the Coulomb interaction on the spatial motion of valence protons in $^8$C in comparison with the valence neutrons in $^8$He.
This is related to the mirror symmetry in two nuclei.
In this study we compare the various radii for $^8$He and $^8$C to investigate the motion of valence nucleons.
Note that the radius of Gamow resonances generally becomes a complex number, 
because the states have the complex amplitudes belonging to the complex energies.
In the numerical results for resonances, their radii show the imaginary parts, which are smaller than the real values.
A similar tendency is confirmed in the squared amplitudes shown in Tables \ref{comp8_1} and \ref{comp8_2}.
From this property, we use the real part of the complex radius to discuss the spatial property of resonances.
The operator forms of the various radii in COSM are given in Ref. \cite{suzuki88}.
The radii of matter ($R_{\rm m}$), proton ($R_{\rm p}$), and neutron ($R_{\rm n}$) have the relation of $AR^2_{\rm m}=ZR^2_{\rm p}+NR^2_{\rm n}$ with proton number $Z$ and neutron number $N$.

\begin{table}[t]  
\caption{Radius of matter ($R_{\rm m}$), proton ($R_{\rm p}$), and neutron ($R_{\rm n}$) of the $0^+_{1,2}$ states in $^8$He and $^8$C in units of fm.}
\label{tab:radius8}
\centering
\begin{tabular}{c|cc|cc}
\noalign{\hrule height 0.5pt}
                        & $^8$He($0^+_1$) & $^8$C($0^+_1$)       & $^8$He($0^+_2$) & $^8$C($0^+_2$)  \\
\noalign{\hrule height 0.5pt}
$R_{\rm m}$             & 2.52            & $2.81-0.08{\it i}$   & 7.56 + 2.04${\it i}$  & 4.87 + 0.13${\it i}$  \\
$R_{\rm p}$             & 1.80            & $3.06-0.10{\it i}$   & 3.15 + 0.69${\it i}$  & 5.46 + 0.15${\it i}$  \\
$R_{\rm n}$             & 2.72            & $1.90-0.01{\it i}$   & 8.53 + 2.32${\it i}$  & 2.36 + 0.05${\it i}$  \\
\noalign{\hrule height 0.5pt}
\end{tabular}
\end{table}

We show the results of the radii for $^8$He and $^8$C in Table \ref{tab:radius8} for the matter, proton, and neutron parts.
For the $0^+_1$ states, the matter radius of $^8$C is larger than the value of $^8$He by about 12\% in the real part.
For $0^+_2$ states, all the radii become complex numbers because of the property of resonances.
Their imaginary parts are shown to be large in $^8$He($0^+_2$), but still smaller than the real parts.
The values in $^8$C have small imaginary parts and the bound-state property can survive for this state.
In two nuclei, various radii for the $0^+_2$ states become larger than those for their ground $0^+_1$ states.
This trend represents the spatial extension of valence nucleons occurring in the $0^+_2$ states.

\begin{figure}[t]
\centering
\includegraphics[width=7cm,clip]{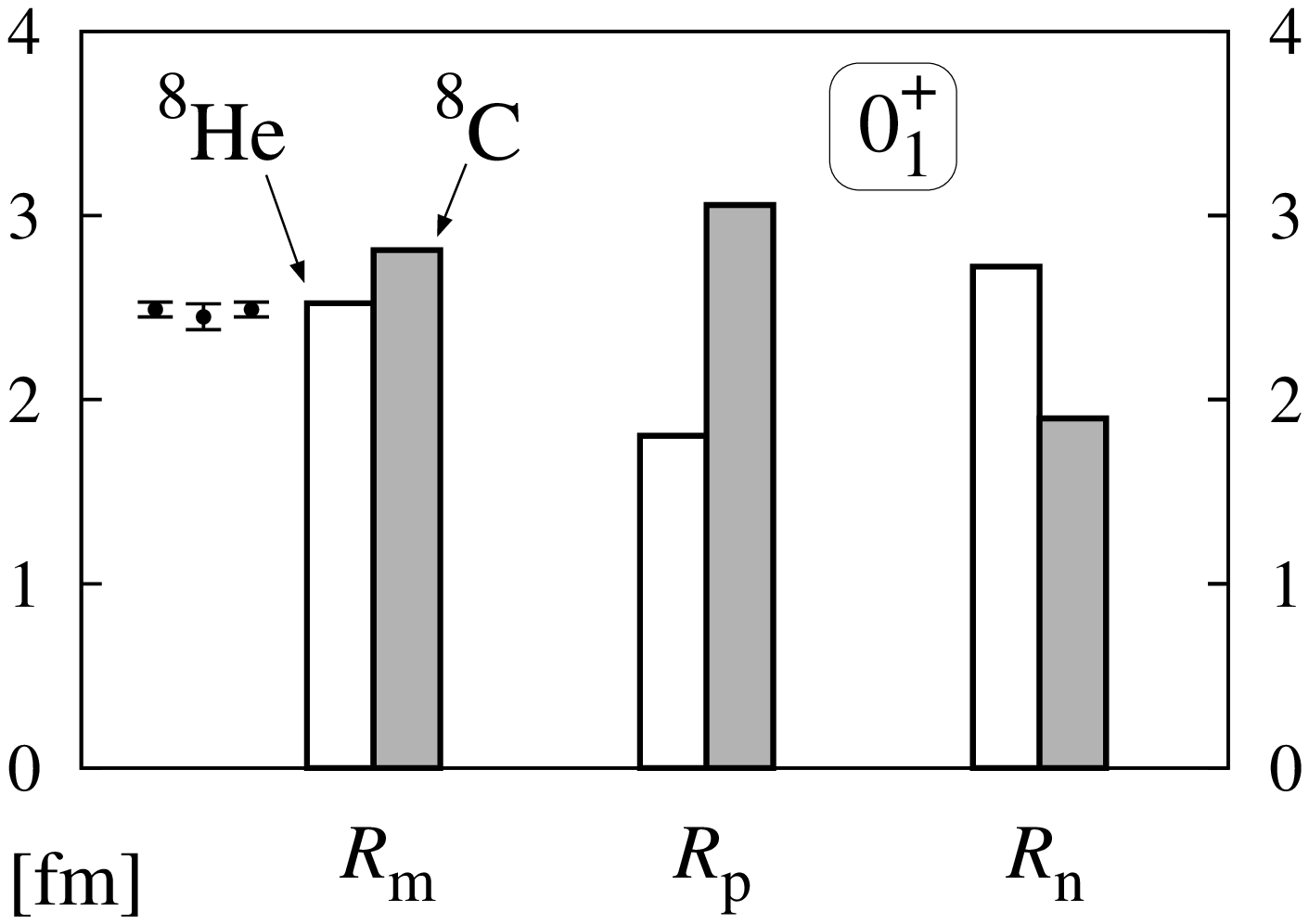}\hspace*{0.9cm}
\includegraphics[width=7cm,clip]{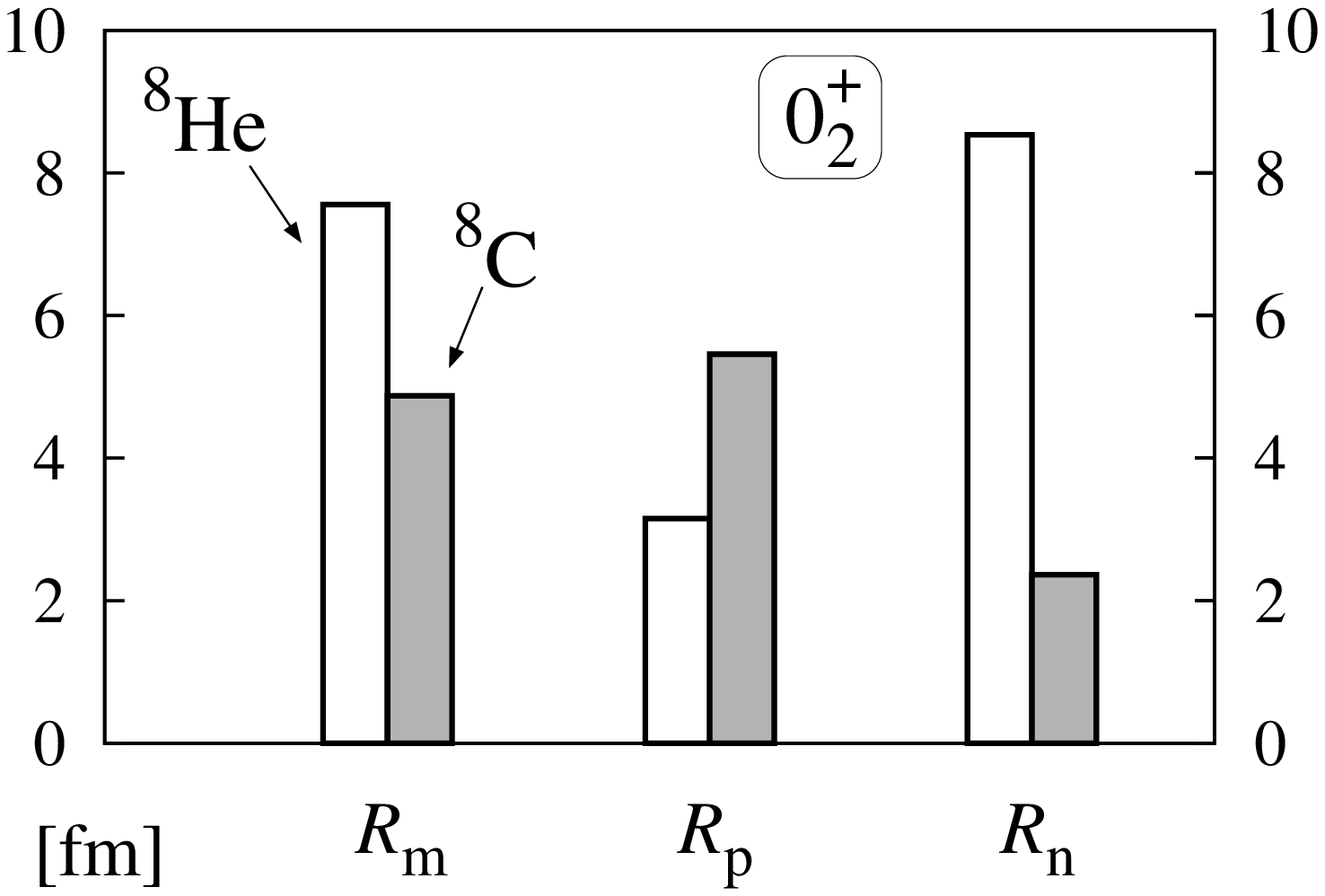}
\caption{Real parts of matter ($R_{\rm m}$), proton  ($R_{\rm p}$), and neutron  ($R_{\rm n}$) radii of $^8$He and $^8$C in units of fm.
Left: Ground states. Right: $0^+_2$ states.  
The circles with error bars are experimental values of the matter radius of $^8$He \cite{tanihata92,alkazov97,kiselev05}.}
\label{fig:radius8}
\end{figure}

It is found that the matter radius of $^8$C is smaller than the value of $^8$He for the $0^+_2$ states.
In particular, the proton radius of $^8$C is found to be smaller than the neutron radius of $^8$He from a mirror relation.
This is an opposite relation to that for the ground $0^+_1$ states of $^8$He and $^8$C.
To make clear this difference, we plot the real parts of the matter, proton, and neutron radii of $0^+_{1.2}$ in Fig. \ref{fig:radius8}.
For $^8$He($0^+_2$), the large matter radius is due to the large neutron radius. 
For  $^8$C($0^+_2$), the large matter radius comes from the large proton radius, which is smaller than the neutron radius of $^8$He.

We conclude that the spatial relation between $^8$He and $^8$C depends on the states, 
and this property is understood from the Coulomb interaction.
The Coulomb interaction acts repulsively and shifts the energy of $^8$C upward from the case of $^8$He
as seen in Fig. \ref{fig:ene_COSM}.
In the ground state of $^8$C, this repulsion increases the distances between $\alpha$ and a valence proton and between valence protons.
In addition, the Coulomb interaction makes the barrier above the particle threshold in $^8$C, and 
the $0^+_2$ resonance is influenced by this barrier, which prevents the four valence protons in $^8$C from extending spatially.  
In $^8$He, there is no Coulomb barrier for valence neutrons and the neutrons in the resonances can be extended to a large distance. 
This effect of the Coulomb interaction makes the radius of $^8$C($0^+_2$) smaller than the value of $^8$He ($0^+_2$).

\section{Complex-scaled Lippmann--Schwinger equation}\label{sec:CSLS}
~~~
In unstable nuclei, separation energies of valence nucleons are much smaller than for stable nuclei by one order.
Due to this weakly bound property, unstable nuclei are broken up in a low-excitation energy.
Utilizing this property, the breakup reactions become important tools experimentally to explore the exotic properties of unstable nuclei \cite{Ta13}.
Theoretically, a reliable description of the many-body scattering states of weakly bound systems is required.

Recently, several theoretical methods have been proposed to solve many-body scattering problems \cite{efros07,Ca14}.
We have also developed a method to describe the many-body scattering states in CSM, called the ``complex-scaled solutions of the Lippmann--Schwinger equation'' (CSLS equation)~\cite{Ki09,Ki10,Ki11,Ki13}.
In the CSLS equation, scattering states are described in the formal solutions of the Lippmann--Schwinger equation using the CSGF.
This CSGF given in Eq. (\ref{eq-2-3-1}) is automatically imposed to satisfy the correct boundary conditions of many-body scattering states using the complex-scaled eigenstates.
We have shown that the CSGF gives the precise CLD of the two-body scattering states \cite{suzuki05,suzuki08} and 
works consistently to obtain the three-body CLD \cite{My14}.
We describe the many-body scattering states of unstable nuclei in the CSLS equation,  
and obtain the physical quantities as functions of the energies of the subsystems in a many-body states. 
This gives useful information on the internal correlations of unstable nuclei.
This approach contributes to clarifying the exotic properties of weakly coupled systems, such as the neutron motion in a halo.

In this paper we explain the following two topics related to the three-body scattering states in the CSLS equation:
(i)  the Coulomb breakup reactions of two-neutron halo nuclei, $^6$He and $^{11}$Li;
(ii) the elastic scattering and radiative capture reaction of the $\alpha$+$d$ system.

We comment on the continuum discretized coupled channel (CDCC) approach for breakup reactions \cite{kamimura86}.
In CDCC, the reaction process between the projectile and target nuclei is dynamically solved in a coupled-channel problem.
In CDCC, the continuum states of the projectile nucleus are discretized in a two-body breakup case.
For three-body breakup cases such as the two-neutron halo nuclei, there is an extension of CDCC by applying ECR and the CSGF to the projectile nucleus \cite{matsumoto10}.

\subsection{Three-body Coulomb breakup reactions of halo nuclei}
~~~Coulomb breakup reactions are utilized to understand the excitation properties of two-neutron halo nuclei such as $^6$He and $^{11}$Li~\cite{aumann99,wang02,ieki93,shimoura95,zinser97,nakamura06},
and in particular the dipole responses of the halo structure.
In two-neutron halo nuclei, no binary subsystems have bound states in a core+$n$+$n$ picture, which is called the Borromean property.
We theoretically describe two-neutron halo nuclei in the core+$n$+$n$ three-body model and investigate their Coulomb breakup reactions.
We describe here the three-body scattering states of halo nuclei by using the CSLS equation.
In this review, we focus on the Coulomb breakup reactions of $^6$He and $^{11}$Li.
For $^{11}$Li, we consider the configuration mixing of the $^9$Li core nucleus with particle--hole excitations, 
which is essential to reproduce the halo properties of $^{11}$Li with a large $s$-wave mixing of two neutrons \cite{myo07b}.

In the Coulomb breakup reactions of two-neutron halo nuclei, the asymptotic Hamiltonian $H_0$ is given as
\begin{equation}
H_0 = h_{\rm core} + \sum_{i=1}^3 t_i - T_{\rm c.m.},
\end{equation}
where $h_{\rm core}$ is the internal Hamiltonian of the core nucleus of the $\alpha$ particle for $^6$He and $^{9}$Li for $^{11}$Li.
The kinetic energy operators for each particle and for the center-of-mass part are $t_i$ and $T_{\rm c.m.}$, respectively.
The solution of $H_0$ is given as
\begin{equation}
\Phi_0 (\vc{k},\vc{K}) = \Phi_{\rm gs}^{\rm core} \otimes \phi_0 (\vc{k},\vc{K}),
\label{eq:asy_halo}
\end{equation}
where $\vc{k}$ and $\vc{K}$ indicate the asymptotic momenta with a three-body Jacobi coordinate.
In the asymptotic region, the core nucleus becomes the ground state $\Phi_{\rm gs}^{\rm core}$.
For $^6$He, the ground state of the $\alpha$ core is the $(0s)^4$ closed configuration \cite{Ki09,Ki10}.
For $^{11}$Li, the ground-state wave function of $^9$Li is obtained in the tensor-optimized shell model (TOSM) \cite{Ki13,myo07b,myo08}.
The asymptotic wave function $\phi_0$ for the relative motion part in the core+$n$+$n$ three-body system is given as
\begin{equation}
\phi_0 (\vc{k},\vc{K}) = \frac{1}{(2\pi)^3}\, e^{i\boldsymbol{k}\cdot\boldsymbol{r}+i\boldsymbol{K}\cdot\boldsymbol{R}},
\end{equation}
where the coordinates $\vc{r}$ and $\vc{R}$ are the conjugates to the momenta $\vc{k}$ and $\vc{K}$, respectively.

We construct the Green's function given in Eq.~(\ref{eq-2-3-1}) and three-body ECR in Eq.~(\ref{eq:ECR3})
to describe the scattering states of two-neutron halo nuclei.
In a similar manner to COSM, we obtain the eigenstates $\{\Psi^\theta_\nu\}$ with the index $\nu$ and their eigenvalues $\{E^\theta_\nu\}$ 
by solving the complex-scaled Schr{\"o}dinger equation,
\begin{equation}
H^\theta \Psi^\theta_\nu = E^\theta_\nu \Psi^\theta_\nu.
\label{eq:cs_Scheq}
\end{equation}
For the core+$n$+$n$ three-body system, the total Hamiltonian $H$ is given as
\begin{equation}
H = h_{\rm core} + \sum_{i=1}^3 t_i - T_{\rm c.m.} + \sum_{i=1}^2 V_{{\rm core\mbox{--}}n} (\vc{r}_i) + V_{n{\rm \mbox{--}}n},
\end{equation}
where $V_{{\rm core\mbox{--}}n}$ and $V_{n{\rm \mbox{--}}n}$ are the interactions for core-$n$ and $n$-$n$, respectively.
The coordinate between the core nucleus and the $i$th neutron is given by $\vc{r}_i$.
We remove the Pauli forbidden states occupied by the nucleons in the core nucleus from the core--$n$ relative motion.
For $^6$He we adopt the same Hamiltonian used in COSM and for $^{11}$Li we adopt the same Hamiltonian defined in Refs.\cite{myo07b,myo08}.
We transform the total Hamiltonian $H$ in CSM and solve the complex-scaled eigenvalue problem with the Hamiltonian $H^\theta$ in Eq.~(\ref{eq:cs_Scheq}).

\subsubsection{Coulomb breakup cross sections of $^6$He and $^{11}$Li}
We show the Coulomb breakup cross sections as functions of the excitation energies of $^6$He and $^{11}$Li.
The target nucleus is Pb and the incident energies of the $^6$He and $^{11}$Li projectiles are 240 and 70 MeV/nucleon, respectively.
The cross sections are evaluated using the $E1$ transition strength with the equivalent photon method~\cite{bertulani88},
\begin{equation}
\frac{d^6\sigma}{d\vc{k} d\vc{K}} = \frac{16\pi^3}{9\hbar c}N (E_\gamma) \frac{d^6 B(E1)}{d\vc{k} d\vc{K}},
\label{eq:cs_six}
\end{equation}
where $N(E_\gamma)$ is the virtual photon number with the photon energy $E_\gamma$ that the projectile nucleus absorbs.
The $E1$ transition strength is obtained with  the CSLS solutions as
\begin{equation}
\frac{d^6 B(E1)}{d\vc{k} d\vc{K}} = \frac{1}{2J_0+1} \left|\big\bra \Psi^{(-)} (\vc{k},\vc{K}) || \widehat O(E1) || \Psi_0 \big\ket\right|^2,
\end{equation}
where the operator $\widehat O(E1)$ represents the $E1$ transition.
The ground-state wave function and the total spin are $\Psi_0$ and $J_0$, respectively.

By integrating the differential cross sections in Eq.~(\ref{eq:cs_six}),
we obtain the cross section as functions of the scattering energy $E$ \cite{Ki09,Ki10,Ki13},
\begin{equation}
\frac{d\sigma}{dE} = \iint d\vc{k}\, d\vc{K} \frac{d^6\sigma}{d\vc{k} d\vc{K}}\
\delta\left(E-\frac{\hbar^2k^2}{2\mu}-\frac{\hbar^2K^2}{2M}\right),
\end{equation}
where $\mu$ and $M$ are the reduced masses for the corresponding momenta.

\begin{figure}[tb]
\includegraphics[width=7.0cm,clip]{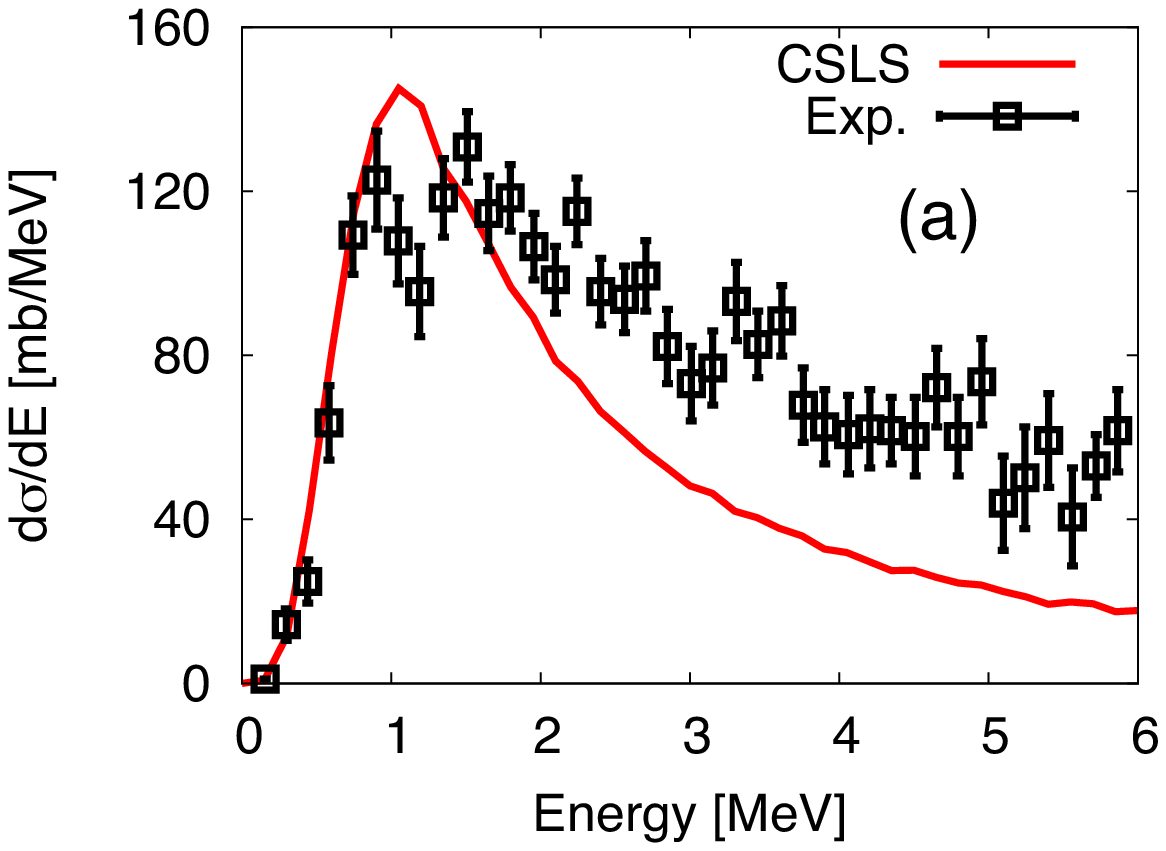}
\hfill
\includegraphics[width=7.0cm,clip]{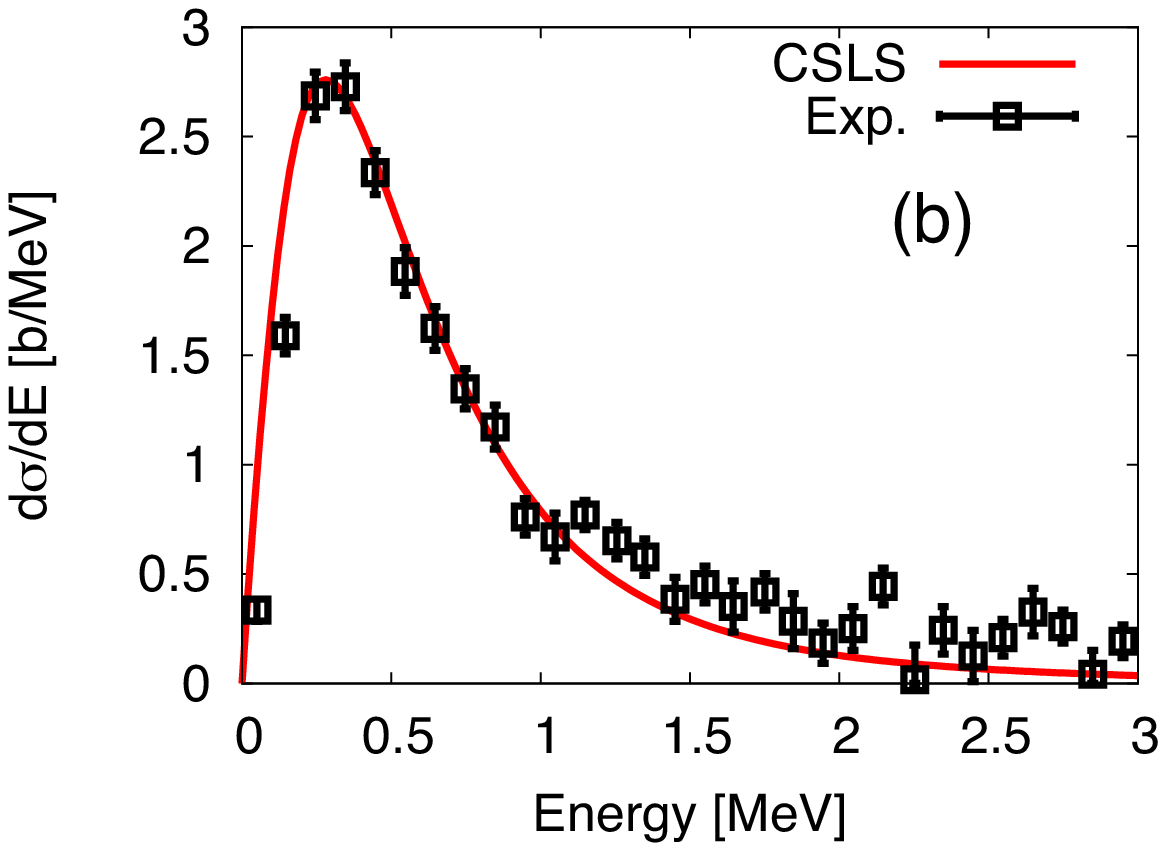}
\caption{Coulomb breakup cross sections measured from the three-body threshold energies of core+$n$+$n$.
Panels (a) and (b) indicate the results for $^6$He and $^{11}$Li, respectively.
The experimental data for $^6$He and $^{11}$Li are taken from Refs.~\cite{aumann99} and \cite{nakamura06}, respectively, shown as open squares with error bars.}
\label{fig:CS}
\end{figure}

In Fig.~\ref{fig:CS}, we show the breakup cross sections of $^6$He and $^{11}$Li measured from the threshold energies of core+$n$+$n$ with the experimental data \cite{aumann99,nakamura06}.
For $^6$He, low-energy enhancement are confirmed around 1 MeV and the cross section decreases gradually with the excitation energy.
The result reproduces the observed cross section \cite{aumann99}, especially at low excitation energy below 2 MeV.
It has been demonstrated that this cross section is dominated by the $^5$He(3/2$^-$)+$n$ decay channel \cite{Ao06,myo01}. 
For $^{11}$Li, the results show good agreement with the experimental data \cite{nakamura06} in the whole energy region.
This cross section also shows a low-energy enhancement around 0.3 MeV and rapidly decreases as the energy increases.

For $^6$He and $^{11}$Li, it has been commonly confirmed that the low-energy enhancements in the cross sections are affected by the strong final-state interactions (FSIs) in the dipole excited states~\cite{Ki10,Ki13}.
This result means that the Coulomb breakup cross sections are much influenced by the final three-body scattering states.
From this property, information on the ground-state structure of halo nuclei is difficult to obtain due to the strong FSI.
We further investigate the invariant mass spectra of the binary subsystems to clarify the mechanisms of the Coulomb breakup reactions in the next section.

Using the three-body ECR of two-neutron halo nuclei given in Eq.~(\ref{eq:ECR3}), we can decompose the breakup cross sections into the components of three-body resonances and two-body and three-body continuum states;
the latter two kinds of continuum states indicate the sequential breakup process and direct breakup, respectively.
We have discussed the amounts of different breakup processes with respect to the total cross sections \cite{myo10,myo03,myo01}.
Note that this decomposition does not correspond to the experimental observables directly.

\subsubsection{Invariant mass spectra for binary subsystems of $^6$He}

We calculate the invariant mass spectra for binary subsystems, 
which is the cross section as functions of the relative energy of the binary subsystems such as core--$n$ and $n$--$n$.
This quantity provides the information on the correlations of the subsystems in the reaction.
Using Eq.~(\ref{eq:cs_six}), we define the invariant mass spectra with the strength distribution obtained in the CSLS equation as
\begin{equation}
\frac{d\sigma}{d\varepsilon} = \iint d\vc{k}\, d\vc{K} \frac{d^6\sigma}{d\vc{k} d\vc{K}}\ \delta \left(\varepsilon - \frac{\hbar^2 k^2}{2\mu}\right),
\label{eq:inv}
\end{equation}
where $\varepsilon$ is the relative energy of the binary subsystem with momentum $\vc{k}$.

\begin{figure}[t]
\centering{\includegraphics[width=7.0cm,clip]{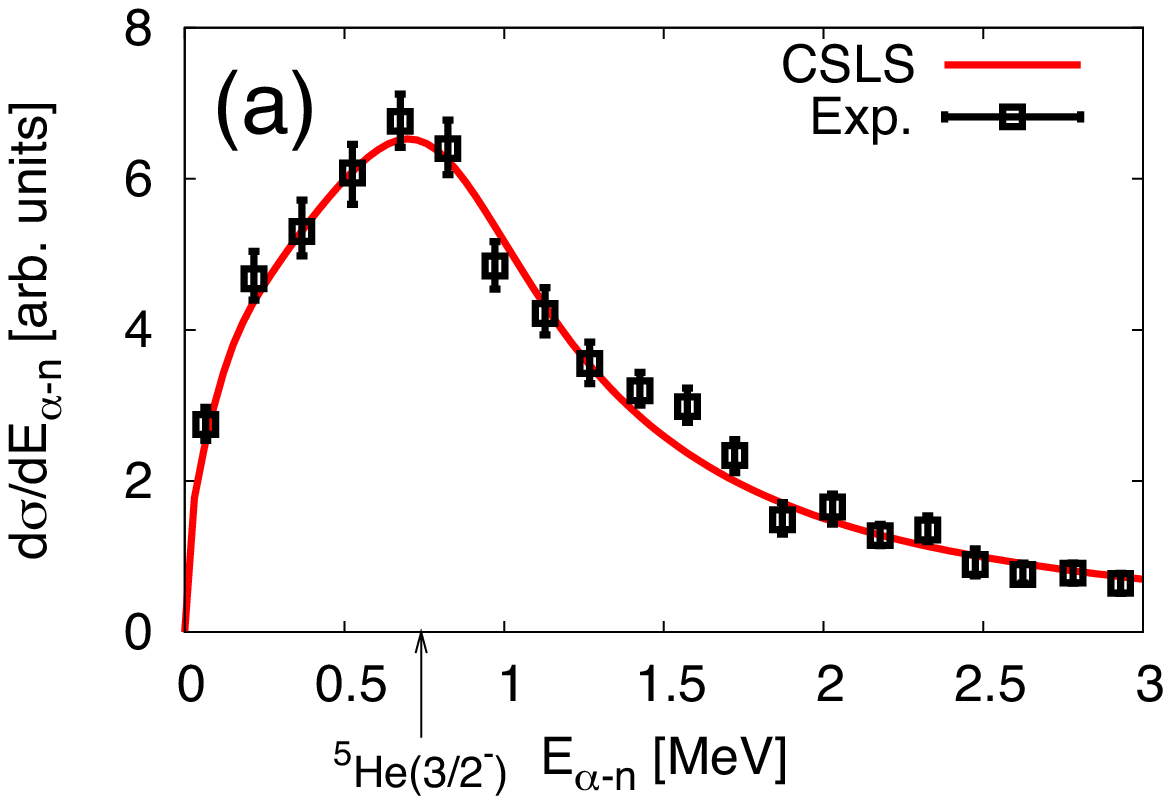}}
\hfill
\centering{\includegraphics[width=7.0cm,clip]{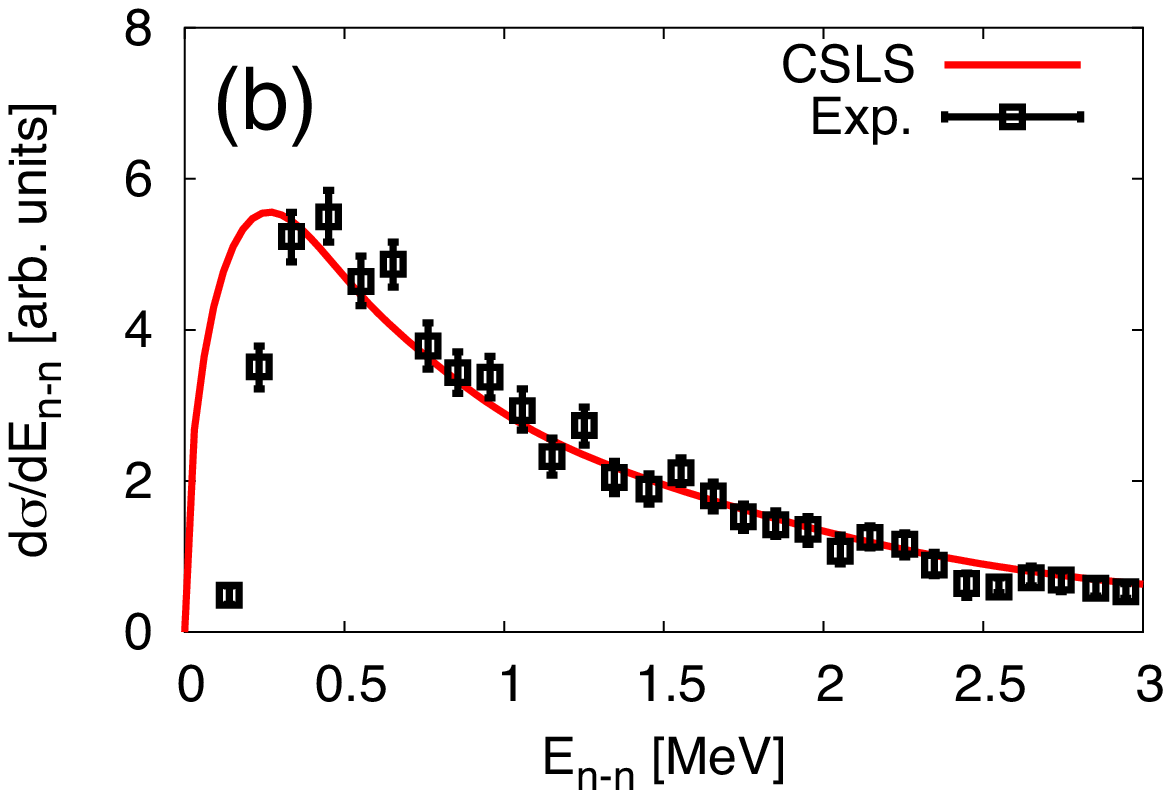}}
\caption{Invariant mass spectra of the Coulomb breakup cross section for $^6$He with arbitrary units. 
Panels (a) and (b) are the strengths as functions of the energies of 
the $\alpha$--$n$ and $n$--$n$ binary subsystems, respectively. 
The experimental data is shown by the open squares \cite{aumann99}.
The arrow in panel (a) shows the $^5$He(3/2$^-$) resonance energy.}
\label{fig:CS_IMS}
\end{figure}

In Fig.~\ref{fig:CS_IMS}, we show the invariant mass spectra of $^6$He as compared with the experimental data \cite{aumann99}.
Panels (a) and (b) are the spectra of the $\alpha$--$n$ and $n$--$n$ subsystems, respectively.
Two spectra show good agreement with the experimental data, which indicates the reliability of the CSLS equation.
For the $\alpha$-$n$ system in Fig.~\ref{fig:CS_IMS} (a), the peak position of the strength agrees with the resonance energy of 0.74 MeV in $^5$He(3/2$^-$).
This result shows the dominance of the sequential breakup process of $^6$He into $\alpha$+$n$+$n$ via the $^5$He+$n$ channel.
For the $n$-$n$ system in Fig.~\ref{fig:CS_IMS} (b), low-energy enhancement is confirmed near the zero energy, which originates from the $n$--$n$ virtual state in the attractive $S$-wave channel.

\subsection{Scattering of $\alpha$ and deuteron in the $\alpha$+$p$+$n$ three-body model}
The deuteron, denoted as $d$, is a weakly bound system with an energy of 2.2 MeV and easily excited inside the nucleus.
In $^6$Li, the threshold energies of the $\alpha$+$d$ and $\alpha$+$p$+$n$ systems are very close, with excitation energies of 1.47 and 3.70 MeV, respectively.
This property suggests that the $\alpha$+$d$ and $\alpha$+$p$+$n$ structures coexist in the low excitation energy region.
In the scatterings associated with $^6$Li, it is important to describe the three-body scattering states with the $\alpha$+$p$+$n$ configurations.
We describe the scattering states of $^6$Li in the CSLS equation, and investigate the $\alpha$+$d$ elastic scattering and the radiative capture reaction of $^2$H($\alpha$,$\gamma$)$^6$Li.
In both reactions, we examine the dynamical effects of $\alpha$+$p$+$n$ three-body structures. 
We investigate the effects of deuteron breakup in $^6$Li and the rearrangement to the $^5$He+$p$ and $^5$Li+$n$ channels on the above reactions \cite{Ki11}.

The scattering states of $^6$Li are described in the $\alpha$+$p$+$n$ three-body model with the CSLS equation.
We define the asymptotic Hamiltonian $H_0$ for the $\alpha$+$d$ system as
\begin{equation}
H_0 = h_d + T_{\rm rel} + V_{\rm Coul} (R),
\end{equation}
where $h_d$ is the internal Hamiltonian of the deuteron with the Argonne V8$^\prime$ realistic $N$--$N$ potential \cite{wiringa02}.
The kinetic energy and the Coulomb interaction between $\alpha$ and $d$ are given as $T_{\rm rel}$ and $V_{\rm Coul}$, respectively.
The relative coordinate between $\alpha$ and $d$ is denoted as $\vc{R}$.
The solution of $H_0$ is expressed as
\begin{equation}
\Phi_0^{\ell J^\pi} (\vc{K},\vc{r},\vc{R}) = \left[\chi^{1^+}_d(\vc{r}) \otimes \phi^\ell_0 (\vc{K},\vc{R})\right]_{J^\pi},
\label{eq:asym_ad}
\end{equation}
where $\ell$ is the relative orbital angular momentum for $\alpha$ ($0^+$) and $d$ ($1^+$), and $J^\pi$ is the total spin and parity.
The relative momentum is given as $\vc{K}$.
The deuteron wave function is $\chi^{1^+}_d(\vc{r})$.
The asymptotic relative wave function $\phi_0^l$ for the $\alpha$~+~$d$ system is defined as
\begin{equation}
\phi_0^\ell (\vc{K},\vc{R}) = (2\ell+1)\, i^\ell\, \frac{F_\ell(\eta,KR)}{KR} \sum_m Y_{\ell m}(\hat{\vc{K}}) Y^*_{\ell m}(\hat{\vc{R}}),
\end{equation}
where $F_\ell$ and $\eta$ are the regular Coulomb wave function and the Sommerfeld parameter, respectively.
Using Eq. (\ref{eq:asym_ad}), we obtain the scattering states of the $\alpha$+$d$ system $\Psi^{(+)}_{\ell J^\pi} (\vc{K})$ in the CSLS equation using Eq.~(\ref{eq-2-6-1}).
We solve the $\alpha$+$p$+$n$ three-body model with Gaussian expansion and complex scaling, and prepare the set of eigenstates $\{\Psi^\theta_\nu\}$.
Using these eigenstates, we describe the $\alpha$+$p$+$n$ three-body components in the Green's function in the CSLS equation.

\subsubsection{Elastic phase shifts of $\alpha$+$d$ scatterings\label{sec:phs}}

\begin{figure}
\begin{minipage}{0.48\textwidth}
\centering\includegraphics[width=7.0cm]{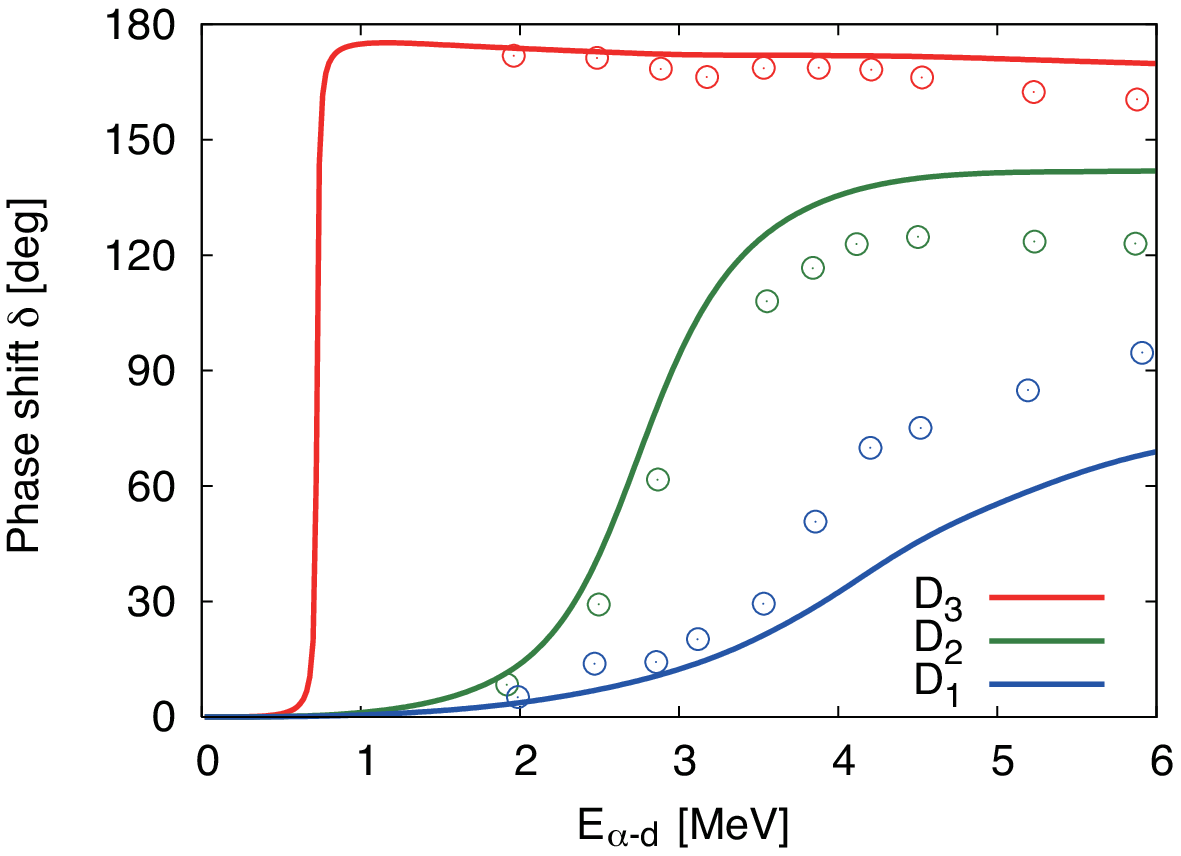}
\caption{\label{fig:phs_6Li}
Elastic $D$-wave phase shifts $D_J$ with spin $J$ for the $\alpha$+$d$ scatterings as functions of the $\alpha$+$d$ relative energies.
The red, green, and blue lines show the calculations for $D_3$, $D_2$, and $D_1$ scattering states, corresponding to the $3^+$, $2^+$, and $1^+$ states, respectively. Experimental results \cite{schmelzbach72,gruebler75} are shown as open circles in the same colors
as the lines.}
\vspace{0.0cm}
\end{minipage}
\hfill
\begin{minipage}{0.48\textwidth}
\centering\includegraphics[width=7.2cm]{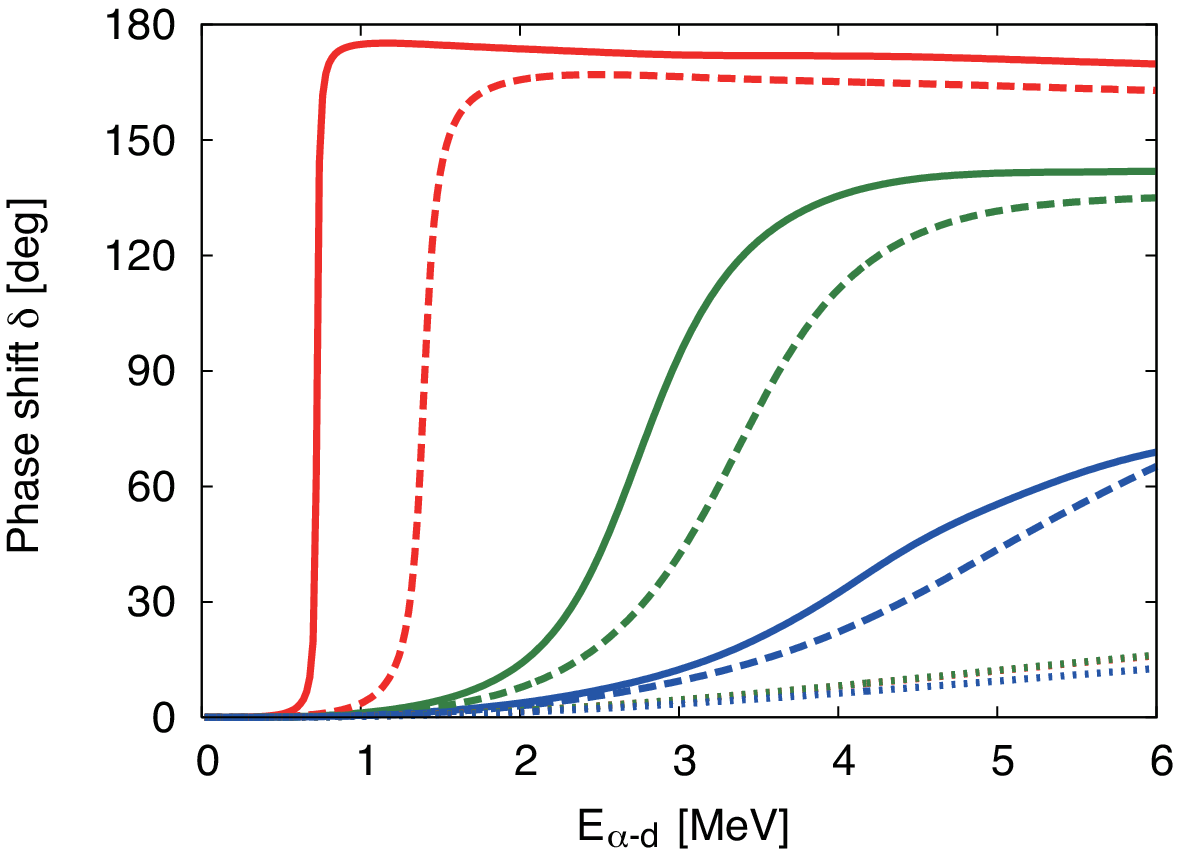}
\caption{\label{fig:phs_comp}
The effects of deuteron breakup and rearrangement on the $\alpha+d$ phase shifts.
The solid lines are the full calculations and are as shown in Fig.~\ref{fig:phs_6Li}. 
The dotted and dashed lines are the Elastic and Breakup calculations, respectively. 
The three dotted lines for the elastic calculation give almost identical solutions.
}\vspace{0.5cm}
\end{minipage}
\end{figure}

In Fig.~\ref{fig:phs_6Li}, we show the $\alpha$+$d$ elastic phase shifts for the relative $D$-wave scatterings with total spin $J$.
The resulting phase shifts reproduce the trend of the experimental data for the $J=1^+$, $2^+$, and $3^+$ states.
This agreement represents the reliability of the CSLS scheme in three-body description of $^6$Li.
Three kinds of the phase shifts show resonance behavior, and
in CSM the resonance energies measured from the $\alpha+d$ threshold and the decay widths ($E_r,\Gamma$)
are obtained as (4.12,~3.60) for $1^+$, (2.83,~0.91) for $2^+$, and (0.72,~0.19) for $3^+$ in units of MeV.

We investigate the effect of a three-body configuration on the $\alpha$+$d$ scatterings.
We focus on two kinds of effects of deuteron breakup and the rearrangement to the $^5$He+$p$ and $^5$Li+$n$ channels.
For this purpose, we show two kinds of calculations in addition to that shown in Fig.~\ref{fig:phs_6Li}.
One is the calculation in which only the elastic channel of $\alpha$+$d$ is considered without the effects of the deuteron breakup and rearrangement, and is named ``Elastic".
The other is the calculation in which only the deuteron breakup is included in the breakup effect, and is named ``Breakup".
From these calculations, we estimate the effects of deuteron breakup and the rearrangement channels on the $\alpha$+$d$ scattering.

For the Elastic case, the $p$+$n$ wave function in $^6$Li is kept as the deuteron during the scattering.
We solve only the relative motion between $\alpha$ and $d$ with the coordinate $\vc{R}$ and obtain the set of new eigenstates $\{\Psi^\theta_{\nu,{\rm el}}\}$ of $^6$Li.
For the Breakup case, we allow excitations of the $p$-$n$ relative motion.
First, we prepare the eigenstates of the $p$+$n$ system with the Gaussian expansion, which involves the ground and excited states.
Next, we solve the coupled-channel problem on the relative motion between $\alpha$ and the set of the $p+n$ states, 
and obtain another set of eigenstates $\{\Psi^\theta_{\nu,{\rm br}}\}$ for $^6$Li .
Using the different sets of eigenstates $\{\Psi^\theta_{\nu}\}$, we obtain two different ECRs for the Elastic and Breakup calculations of $^6$Li.

We show the results in Fig.~\ref{fig:phs_comp}.
In the Elastic case with dotted lines, the deuteron is kept in the ground state and all the phase shifts exhibit no structure.
In the Breakup case with dashed lines, the calculations show resonance behaviors for each state.
These results indicate that the deuteron breakup is very important in $\alpha$+$d$ scattering, as suggested in Ref.~\cite{kamimura86}.
Note that the positions of the resonance energies are higher than in the full calculations.
This difference indicates the rearrangement effect, which shifts the resonance energies down by about 0.5 MeV.
In the present analysis the deuteron breakup is found to have a significant role on the resonance description of $^6$Li in the $\alpha$+$d$ scattering, 
while the rearrangement effects give a small contribution to the phase shift.

\subsubsection{Radiative capture cross section for $^6$Li}

\begin{figure}[t]
\begin{minipage}{0.48\textwidth}
\centering\includegraphics[width=7.0cm]{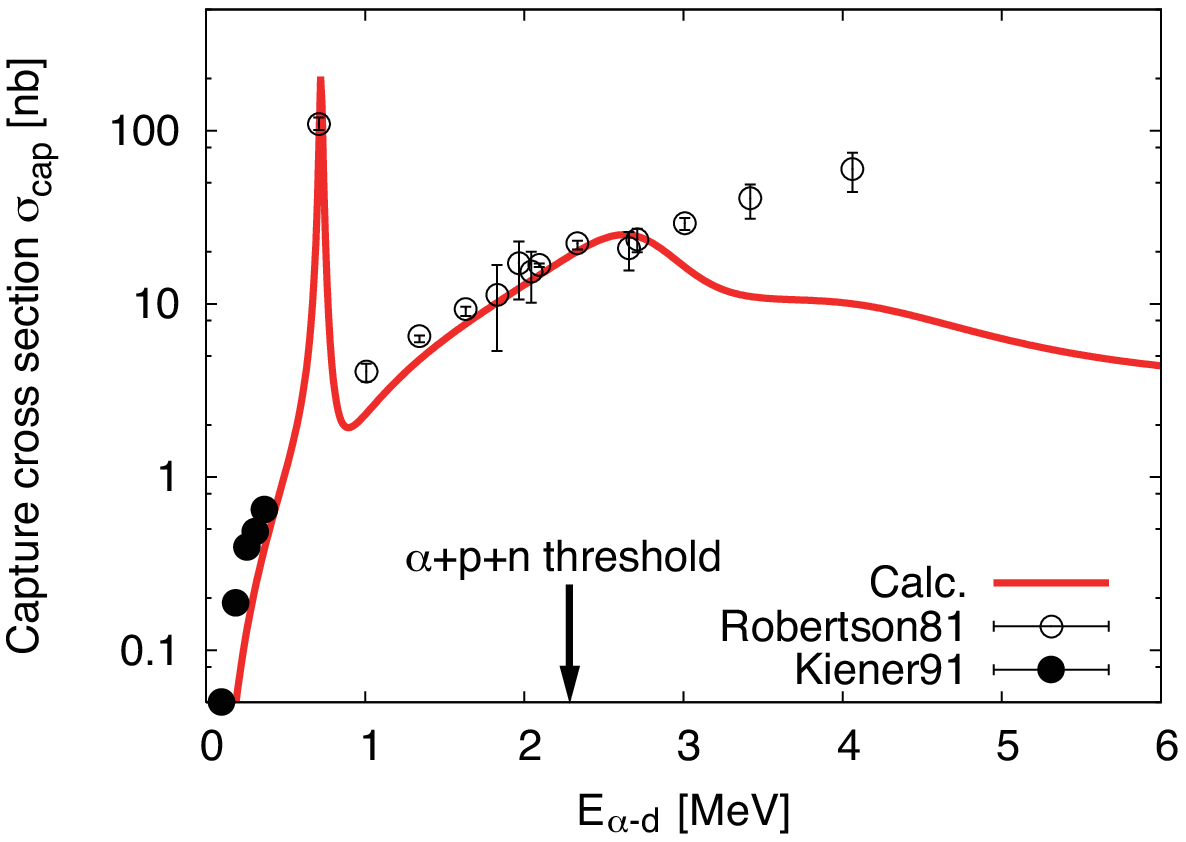}
\caption{\label{fig:pcs_6Li}
The radiative capture cross section of $^2$H$(\alpha,\gamma)^6$Li is shown with red lines.
The open and solid circles with error bars are the experimental data from Refs. \cite{robertson81} and \cite{kiener91}, respectively. 
}\vspace*{0.8cm}
\end{minipage}
\hfill
\begin{minipage}{0.48\textwidth}
\centering\includegraphics[width=7.0cm]{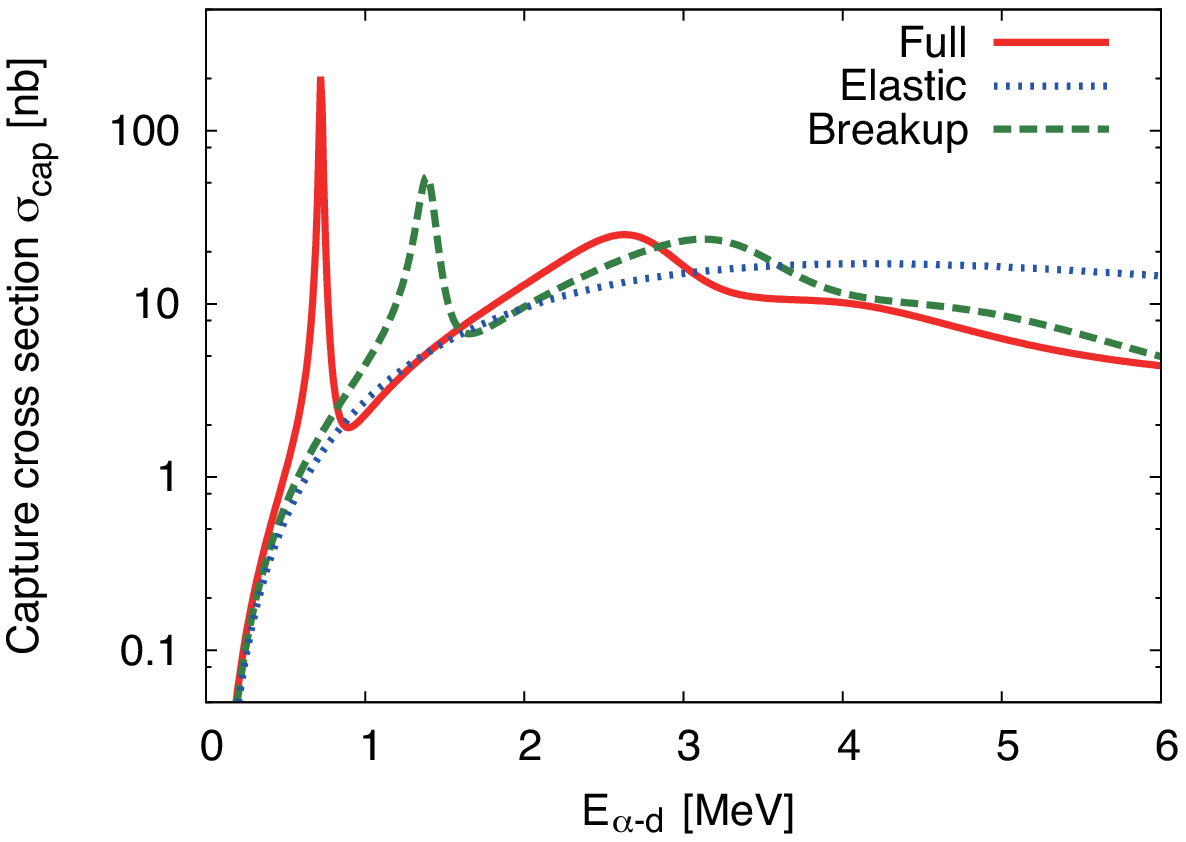}
\caption{\label{fig:pcs_comp}
The effects of deuteron breakup and rearrangement on the radiative capture cross section of $^6$Li. 
The red solid line is the same as in Fig.~\ref{fig:pcs_6Li}. 
The blue dotted and green dashed lines are the results of the Elastic and Breakup cases, respectively.
}
\end{minipage}
\end{figure}

We now discuss the $\alpha$+$p$+$n$ three-body effect on the radiative capture reactions of $^6$Li.
We calculate the radiative capture cross section of $^2$H$(\alpha,\gamma)^6$Li, $\sigma_{\rm cap}$, in the relation
\begin{equation}
\sigma_{\rm dis} (E) = \frac{2(2J_{\rm gs}+1)}{(2J_\alpha+1)(2J_d+1)}\frac{k_\gamma^2}{K^2}\, \sigma_{\rm cap} (E),
\label{eq:radcap}
\end{equation}
where $J_{\rm gs}$, $J_\alpha$ and $J_d$ are the spins of the $^6$Li ground state ($1^+$), $\alpha$, and deuteron, respectively.
The wave number of the emitted photon is $k_\gamma = E_\gamma/\hbar c$.
The photodisintegration cross section $\sigma_{\rm dis}$ is obtained using
\begin{equation}
\sigma_{\rm dis} (E) = \frac{4\pi^3}{75} \left(\frac{E_\gamma}{\hbar c}\right)^3 \int d\vc{K}\, \frac{d^3 B(E2)}{d\vc{K}}\, \delta \left( E - \frac{\hbar^2 K^2}{2M} + \varepsilon_d \right),
\label{eq:disint}
\end{equation}
where $M$ is the reduced mass for the momentum $\vc{K}$ and $\varepsilon_d$ is the binding energy of the deuteron.
The photon energy is defined as $E_\gamma=E+\varepsilon_{\rm gs}$, and $\varepsilon_{\rm gs}$ is the binding energy of the $^6$Li ground state with respect to the $\alpha$+$p$+$n$ threshold energy.
In this calculation, we consider the contribution of the dominant $E2$ transition.
The $E2$ transition strength is defined as
\begin{equation}
\frac{d^3B(E2)}{d\vc{K}} = \frac{1}{2J_{\rm gs}+1} \left|\bra \Psi^{(-)}_{\ell J^\pi} (\vc{K}) || \widehat O(E2) || \Psi_0 \ket\right|^2,
\label{eq:disE2}
\end{equation}
where the operator $O(E2)$ represents the $E2$ transition and $\Psi_0$ is the wave function of the $^6$Li ground state.

In Fig.~\ref{fig:pcs_6Li}, we show the radiative capture cross section of $^2$H$(\alpha,\gamma)^6$Li.
The calculation shows a good agreement with experiment below the $\alpha$+$p$+$n$ threshold energy.
Above the energy of $E_{\alpha\mbox{-}d} = 3$ MeV, the present result underestimates three of the experimental data points.
One possibility to improve this underestimation is to include higher-order transitions, such as the $M1$ and multi-step transitions.

We estimate the $\alpha$+$p$+$n$ three-body effect on the cross section.
For this purpose, we perform the same analysis as for the $\alpha+d$ scattering.
We show the results in Fig.~\ref{fig:pcs_comp}.
Similar to the phase shift case, the Elastic result shows a structureless distribution.
The deuteron breakup result (Breakup) shows peaks and bumps, the energies of which are the $3^+_1$, $2^+_1$, and $1^+_2$ resonances, 
although these energies are slightly higher than the full three-body calculations of $^6$Li.
This difference is explained via the $^5$He+$p$ and $^5$Li+$n$ rearrangement channels in a similar situation to the phase shifts.

\section{Summary and perspective}
We have explained the frameworks of the complex scaling method (CSM) to study many-body resonances and continuum states, and presented applications to many-body nuclear systems in the recently developed physics of unstable nuclei. CSM has been developed as a very promising method to obtain the resonance energies and decay widths of many-body systems, and here we emphasized that CSM based on the non-Hermitian Hamiltonian provides us with a powerful method to investigate not only the resonance state but also many-body continuum states.

The basic idea of CSM is that bound, resonant, and continuum states satisfying the boundary condition of the asymptotic outgoing waves with an $\exp{(ikr)}$ form can be expressed with square-integrable functions ($L^2$ class) by a complex scaling (a dilation transformation). This means first that the non-Hermiticity of the complex-scaled Hamiltonians is caused by the boundary condition of the outgoing waves for the state space. This idea indicates that bound, resonant and continuum states are obtained simultaneously by solving an eigenvalue problem using a set of appropriate basis functions. The obtained eigensolutions describe a complete set, and the Green's function under the outgoing wave boundary condition is constructed in terms of these eigensolutions in CSM. This Green's function is a spectrum representation consisting of the bound, resonant, and continuum energy states, and gives a kind of projection of the matrix elements associated with complex energy states into an observable quantity defined on the real energy axis. Thus, although the matrix elements of resonant and continuum energy states are complex numbers, the quantities projected on the real axis are real numbers and can be investigated through the decomposition into each resonant state and different kinds of continuum sates.

After explaining the frameworks of CSM, we showed its basic application to the simple two-body systems. An important advantage of CSM is a natural description of many-body systems, and its application to nuclear many-body systems including unstable nuclei was also demonstrated. Most of the states in an unstable nucleus are observed as unbound states due to the weak binding nature of valence nucleons. CSM provides us with a powerful framework to investigate the interesting properties of unstable nuclei.
We presented our recent results fore many-body systems of up to five bodies.

One of the main interests in the many-body resonances is the breakup dynamics associated with correlations between constituents. To see such correlations among constituents,
we have developed the complex-scaled Lippmann--Schwinger equation. The validity of this method was shown for three-body breakup reactions of neutron halo nuclei.
 
Thus, CSM is expected to bring a unified description of the structures and reactions of nuclei, though there are still many problems to be tackled.
Several theories have put forth ab-initio descriptions of resonances developed from the bare 
nucleon--nucleon interaction, such as the few-body method \cite{Ca14,Ho13} and the no-core shell model approach \cite{Ba13}.
Further extension of the many-body resonant and continuum states covering a wide range of mass numbers is the current task in this direction.
The application of CSM can be extended to hadron and strangeness physics.
There are studies of many-body resonances observed in hypernuclei with hyperons such as $^7_\Lambda$He \cite{hiyama15}, and also in mesic nuclei 
consisting of baryons and mesons such as $\bar K NN$ \cite{dote13,dote18}.

In CSM, because resonant states are described with square-integrable functions as well as bound states, we can investigate their structures by analyzing the wave functions directly.
Various kinds of structures of resonant states are expected, as discussed in bound states, in many-body nuclear systems.
An important subject to investigate in nuclear cluster physics is multi-cluster structure,
such as an $\alpha$ linear-chain structure, in resonant states observed around the corresponding multi-cluster threshold energies \cite{Ho12}.
In atomic and molecular physics, two kinds of resonances have been discussed, a shape-type resonance and a Feshbach-type one \cite{Mo11}.

CSM describes resonances with complex energy eigenvalues, the imaginary part of which represents the total decay width. It is important to evaluate the partial decay widths of many-body resonances for each decaying channel, which provide useful information on the decay properties of the many-body resonances, but are not yet available. It is desirable to develop a method for extracting the partial decay widths of the many-body resonances in CSM. 
There is a theoretical development of obtaining the partial decay widths by using the continuity equation based on the time-dependent Schr\"odinger equation \cite{Mo11,Go10}.

In addition to the resonances, the virtual states are a kind of unbound state which often play an important role in nuclear structure around threshold energies, and are difficult to obtain directly in CSM with $\theta<\pi/2$, different from resonance poles.
The halo structure of $^{11}$Li with a large mixing of the $s$-wave component is closely related to the presence of the virtual $s$-wave states in $^{10}$Li near the $^9$Li+$n$ threshold energy, indicating the attractive effect between $^9$Li and $n$ \cite{Ma00}.
It is desirable to develop a theoretical framework that can treat virtual states in many-body systems \cite{Od15,Od17}.         

\section*{Acknowledgement}
The authors would like to acknowledge collaborations with K. Ikeda, S. Aoyama, M. Homma, R. Suzuki, C. Kurokawa, B. G. Giraud, H. Masui, Y. Kikuchi and M. Odsuren for the development of the complex scaling method in the application to the nuclear physics.
The authors would also like to acknowledge the support of JSPS KAKENHI Grant Nos. 25400241, 15K05091, JP17K05430, and JP18K03660. 


%

\def\JL#1#2#3#4{ {{\rm #1}} \textbf{#2}, #3 (#4)}  
\nc{\PPNP}[3]   {\JL{Prog. Part. Nucl. Phys.}{#1}{#2}{#3}.}
\nc{\PTEP}[3]   {\JL{Prog. Theor. Exp. Phys.}{#1}{#2}{#3}.}

\end{document}